\documentclass{article}

\usepackage[utf8]{inputenc}

\usepackage{url}

\usepackage[pdftex]{graphicx}
\usepackage{xspace}
\usepackage[in]{fullpage}
\usepackage{amsmath}
\usepackage{amsfonts}
\usepackage{amssymb}
\usepackage{amsthm}
\usepackage{xargs}   
\usepackage{color}
\usepackage[pdftex,dvipsnames]{xcolor} 
\usepackage[disable]{todonotes} 

\usepackage{bigstrut}
\usepackage{hyperref}

\newcommand{\cjc}[2][1=]{\todo[linecolor=blue,backgroundcolor=blue!25,bordercolor=blue]{CJC:#2}}
\newcommand{\vjt}[2][1=]{\todo[linecolor=green,backgroundcolor=green!25,bordercolor=green]{VJT:#2}}

\newcommand{\commentOut}[1]{}
\newcommand{\HMAC}{\textit{HMAC}}

\newcommand{\DE}[1]{}
\newcommand{\etal}{et al.\xspace}

\title{Options for encoding names for data linking at the Australian Bureau of Statistics}

\date{\today}
\author{Chris Culnane, Benjamin I.~P. Rubinstein, Vanessa Teague \\
University of Melbourne \\
{\tt \{vjteague,  benjamin.rubinstein, christopher.culnane\}@unimelb.edu.au } }

\begin{document}
\maketitle
\vspace{-1cm}
\section*{Background and scope}
Publicly, ABS has said it would use a cryptographic hash function to convert names collected in the 2016 Census of Population and Housing into an unrecognisable value in a way that is not reversible.  In 2016, the ABS engaged the University of Melbourne to provide expert advice on cryptographic hash functions to meet this objective.\footnote{University of Melbourne Research Contract 85449779}
After receiving a draft of this report, ABS conducted a further assessment of Options 2 and 3, which will be published on their website. 

\section*{Summary}
For complex unit-record level data, including Census data, auxiliary data can be often be used to link individual records, even without names.  This is the basis of ABS’s existing bronze linking.  
This means that records can probably be re-identified without the encoded name anyway.  Protection against re-identification depends on good processes within ABS.

The undertaking on the encoding of names should therefore be considered in the full context of auxiliary data and ABS processes.  There are several reasonable interpretations:
\begin{enumerate}  
\item	That the encoding cannot be reversed except with a secret key held by ABS.  This is the property achieved by encryption (Option 1), if properly implemented;
\item That the encoding, taken alone without auxiliary data, cannot be reversed to a single value.  This is the property achieved by lossy encoding (Option 2), if properly implemented;
\item That the encoding doesn't make re-identification easier, or increase the number of records that can be re-identified, except with a secret key held by ABS.  This is the property achieved by  HMAC-based linkage key derivation using subsets of attributes (Option 3), if properly implemented.
\end{enumerate}

Each option has advantages and disadvantages.  In this report, we explain and compare the privacy and accuracy guarantees of five different possible approaches.  Options~4 and~5 investigate more sophisticated options for future data linking, though are probably not feasible for this year.  We also explain how some commonly-advocated techniques can be reversed, and hence should not be used.

We examine the mathematical properties of each technique in order to explain what the assumptions on procedural protections are, for example whether there are keys that must be kept secret and whether the data remains re-identifiable.  The security guarantees therefore depend on ABS processes for protecting whatever data remains sensitive, such as re-identifiable linked data.  Our aim is to explain clearly what must be protected, for each proposed encoding method.  We understand that ABS will be implementing additional IT security measures and processes such as encryption at rest and access control, although these are not within the scope of this report.

\newpage

\tableofcontents
\newpage
\section{Introduction}
Cryptographers take great care in defining
\begin{itemize}
    \item the abilities of an \emph{attacker} and
    \item the \emph{security guarantees} of a protocol.
\end{itemize}

A cryptographic primitive such as hashing, encryption, or digital signatures might provide certain guarantees against a particular kind of attacker, but might not be secure against a stronger attacker or in a different context.  For example, digital signatures guarantee the integrity of data (assuming the key is secure) but do not provide privacy; encryption schemes that were secure 30 years ago can be broken using modern cloud computing. 

Our first step is to model carefully the attacker ABS needs to defend against.  A technique that defends against trusted parties doesn't necessarily defend against a motivated external attacker.  For example, writing ``confidential'' on the outside of an envelope is an effective way of telling well-behaved people not to read the contents---it is not an effective way of keeping the contents secure from an adversary who wants to snoop.  Much of the Privacy Preserving Record Linkage literature is oriented to defending against well-behaved researchers who don't actively try to reverse protections, like the people who don't open ``confidential'' envelopes. 
\DE{This is the scenario ABS is interested in.  VT: Ultimately this is  your choice, though our recommendation is to consider a motivated attacker.  I hope the report is clear about which constructions defend against that.} 
It is very important not to confuse this level of protection with something that cannot be reversed even by a motivated attacker.  

The world is changing.  We see more active and sophisticated attacks against government infrastructure.  Espionage is conducted by well-funded nation-state attackers against government and corporate databases.  The (allegedly) Russian attacks on the US Democratic National Committee emails were widely publicised.  Less well known but even more devastating was an intrusion into the US Office of Personnel Management~\cite{finklea2015cyber},  blamed on China.  Exfiltrated data contained details about military and intelligence personnel, including information given for security clearances.  
In a separate incident, an employee of the US National Security Agency was reported to have accidentally exposed their collection of powerful hacking tools~\cite{NSAHack}.  

It is also important to consider re-identification of individuals based on the data itself---birthdates and suburbs could uniquely identify many households.  Although this report focuses on ways to protect names (and addresses), any solution should be carefully aligned with secure methods for protecting the rest of the data.


The risk for ABS is that data could be deliberately stolen or accidentally exposed---and would then be subject to deliberate attack.  The key is to   
assess the security of proposals given a clearly defined and accurate attacker model.

\subsection{Overview of the technical challenges}
There are two quite separate technical problems:

\begin{description}
    \item[The linking problem] Maximising the accuracy of linking, both for reducing false matches and failures to find a match.  The same person might have two different-looking names, due to typos, reading errors, changes of address, {\it etc.}  Any solution needs to be robust against these small changes---this is called ``fuzzy matching'' or ``probabilistic matching.'' 
    \item[The cryptographic problem] Clarifying the assumptions behind techniques for keeping data secure. 
    The key is to be explicit about what the security assumptions are so that ABS can make sure they are valid in practice.  \vjt{ABS asked for rewording of this---I've removed the line about linking implying recovery and just written about the explicit assumptions, but still quite different from what they asked for.  Recheck.}
\end{description}

Each of these problems is challenging on its own and presents tradeoffs among different requirements.  For example, solutions to the cryptographic problem involve a tradeoff among the strength of the security guarantees, the computational burden, and the complexity of the cryptographic protocol.  Linking policies must slide a scale between false positives and false negatives.   The combination of the cryptographic problem with the linking problem is particularly difficult.  Along with the tradeoffs within each issue, the security requirements are hard to reconcile with a rich set of linking policies.

Some key themes and questions:
\begin{itemize}
\item the separation of roles, and whether those separations could be implemented with cryptography, or by the access control mechanisms to be put in place by the Statistical Business Transformation Project (SBTP),
\item the protection of other data, rather than only names and addresses, perhaps using encryption,
\item the possibility to link using other data (not only names). 
\end{itemize}

Many published schemes combine techniques for linking with some method of securing data, but the two are conceptually separate.  For example, the technique by Schell {\it et al.}  \cite{schnell2009privacy} combines a preprocessing stage of extracting n-grams from names to get nearby or likely alternatives, with a Bloom filter that finds exact matches.  (This scheme is also mentioned in various surveys \cite{vatsalan2013taxonomy}.) The overall scheme  deals with fuzzy matching, but the two techniques could be analysed and re-used separately.  For example, the same preprocessing stage could be used before a more secure way of making exact matches.  It is the n-gram treatment, not the Bloom filter, that allows for fuzzy matching.

The choice of method for securing the data does impact on the possibilities for fuzzy matching.  Encrypted data can be decrypted (with the right key) so that fuzzy matching can be performed on the decrypted data.  Cryptographic hashing will check for exact matches only---a small change in the input causes a very different output of the hash.  Some forms of encryption (homomorphic encryption) allow for certain kinds of edits to encrypted data.  These schemes are promising for the future, but probably too computationally expensive for use this year.

Public attention before the census focused strongly on the retention of names.  However, there are many other important aspects to a thorough protection of privacy, since a person's date of birth, place of residence and other data could probably be used to identify them in many cases, even without the name.  This is beneficial for linking, but it also represents a risk to privacy in the event that the dataset is leaked or attacked.  

ABS has an entire governance structure, suite of legislation, policies and practices for managing risks associated with the confidentiality of data releases to external users. All ABS staff sign various legal undertakings upon joining the ABS and at regular intervals of time. The Acts under which ABS operates require them to protect the confidentiality of data when released and the legal undertakings signed by ABS staff give an assurance that ABS staff will abide by the ABS Acts as well other relevant Commonwealth Acts.  However, this depends on both good intentions and sound engineering.  Not everyone who wants to keep data secure  understands the complex interaction of assumptions and protocols needed for security.  \vjt{Add ref to the Qld scientists who were barred from using ABS.} 

For example, the recent open publication of re-identifiable MBS-PBS records \cite{culnane2017health} could be attributed to a mismatch between the \emph{assumptions} behind the mathematical protections and the \emph{access protections}, which were non-existent.  The mathematical techniques used for that dataset might have been sufficient for a secure research environment, but were not sufficient for open publication.


The purpose of this document is to clarify the assumptions of the cryptographic protocols for protecting data.  Then ABS can ensure that the security guarantees of processes at ABS match those assumptions.  For example, if a particular method relies on keeping a decryption key secret from the adversary, then ABS must have processes in place for protecting that key.  If the data itself is re-identifiable (and detailed unit-record level data generally is) then the data itself must be protected.  If security relies on the attacker never having access to the Librarian or Linker, then those computers must be very carefully isolated.

The following suggestions consider the security of the whole process, with an effort to remain consistent with ABS's existing linking structure.  


Some initial suggestions on general security:
\begin{description}
\item[Encrypt the analysis items] with a key not known to anyone in the linking process.
\item[Shuffle the output order of the lists.] Otherwise names and data might be recovered simply using order.
\end{description}

We concentrate on privacy, not integrity---an attacker trying to modify the data will generally not be detected or prevented. 

None of the techniques in this report are secure against a compromised Linker.  We assume that the dataset to be linked arrives in plaintext, so the linker has the information necessary to link by definition.  In the future, it would be better to transmit incoming datasets in encrypted form.  Then it might be possible to link without the Linker observing any plaintext records, so even a compromised linker could not reverse names.  This is the main advantage of Options 4 and 5 (Sections~\ref{individual_ids} and~\ref{hom}), which are promising directions for the future though they are probably too complex to implement this year.  

\subsection{The options}
We have five options, each with some variations.  
Since the data could be re-identifiable with some auxiliary information anyway, even without the name,
we concentrate on clarifying what extra information or access is required to perform (authorised or unauthorised) linking or reversing.  The choice of a good solution can then focus on which assumptions are valid in the ABS environment and which controls can be put in place.

The options
\begin{enumerate}
\item Encryption. Encrypt the names with the Linker’s key; keep the key carefully secured.  Section~\ref{sec:encrypt}.
\item Lossy encoding for names.  Section~\ref{sec:lossy}.
\item HMAC-based linkage key derivation using subsets of attributes, like UK ONS. Keep the key carefully secured.  Section~\ref{hmac_linking_keys}.
\item Assign each person a unique ID before linking. Section~\ref{individual_ids}.
\item Homomorphic encryption / secure computation. Section~\ref{hom}.
\end{enumerate}

In Section~\ref{sec:BloomFilters} we explain why Bloom Filters are not a privacy-preserving data structure, and conduct an empirical investigation of the linking quality of some of the constructions in the literature, including the combination of n-grams and Bloom filters.  A broader literature review is included in Appendix~\ref{sec:litreview}.

Before describing the options, we describe background cryptography in Section~\ref{sec:scene}, then ABS's security and functionality requirements in Section~\ref{sec:requirements}.  We first explain why names that can be individually linked can also be reversed.

\subsection{Why names that can be individually linked from plaintext can also be reversed}
\label{sec:presinfo}

\subsubsection{Linking by guessing all possible names}  \label{DictionaryAttack}
It is not possible to give each name a unique encoding that allows one-to-one matching with plaintext names, but is not reversible.

To see why, suppose the ABS holds a database that includes a unique encoding for each name.  There must be some process for matching those encodings with the names in a new, incoming database.  This process must include, somehow, comparing a plaintext name with an encoded one to see whether they match.  But that process clearly implies the capacity to link any individual name to its encoding---an attacker could run through the ABS database checking each encoded name against ``Rubinstein'' or ``Teague'' until there was a match.   Alternatively, for a given encoded name, the attacker could run through a list of all possible names until there was a match. This allows the attacker to find the name that matches any chosen encoding, regardless of whether that name actually appears in an incoming database.

This is not a question of hashing vs encryption, but a fundamental limit of the information that is retained.  Whenever there is a capacity to do individual linking by name, that capacity also permits the encoding to be reversed.  This is true for any way of preserving the name information, including hashing, encryption, HMAC, or simply replacing the names with a random ID and using a lookup table.

Several options exist for putting procedural and mathematical controls in the way of unauthorised access, while still retaining the ability to link. For example, a cryptographic decryption key could be required for linking---this key could be carefully protected or shared among multiple trustworthy people within ABS, using secret sharing so that multiple people had to work together to decrypt the data.  However, the fundamental limit remains: if the key allows linking, it can also be used to recover names.  So ``not reversible'' is an impossibly high bar to set while still being able to perform exact matching to a unique encoded name. 

One way around this is to use lossy encoding, meaning that several different names are mapped to the same encoding so information is truly lost. The ABS already has a technique for doing this, in which names are effectively assigned to bins, creating a level of indistinguishability between names within the same bin. In such systems there is a reduced amount of information preserved, though some still remains. The amount preserved is proportional to the size of the bins and the frequency of the names within those bins. Such approaches can reduce total information held, but at the cost of accuracy of record linking.  In particular it would prevent exact matching of names. It also has some other important security limitations, because it reduces the attacker's information only as much as it reduces the information available for authorised linking---see Section~\ref{sec:lossy}.

\section{Background on Cryptography and possible attacks}\label{sec:scene}
This section presents very brief informal definitions of cryptographic primitives. More formal definitions can be found in~\cite{daboShoupCryptoBook}.  We then explain some known attacks applicable in this setting and describe which methods defend against them.

A \emph{cryptographic hash function} $H$ takes a message $m$ and outputs a hash $h$.  It should be infeasible to recover $m$ given $h$ \emph{if $m$ was randomly chosen from the full input space of $H$.}

A \emph{message authentication code (MAC)} is similar to a hash function, but requires a secret key $k$ to compute the hash.  It should be infeasible to compute the correct hash without knowledge of $k$.

An \emph{HMAC} \cite{krawczyk1997hmac} is a particular kind of MAC.  Under certain assumptions, an HMAC's output cannot be distinguished from random without knowing the key~\cite{bellare2006new}.  

A \emph{secret-key encryption function} takes a key $k$ and message $m$ (called the ``plaintext'')  and produces a ciphertext $c$.  The message $m$ can be recovered from $c$ using $k$.  (This is called ``decrypting''.)   Decrypting should be infeasible without $k$.

A \emph{public-key encryption function} uses two different keys: a \emph{public key} for encrypting the message $m$ and a \emph{private key} for decrypting the ciphertext $c$.  The public key is made public, but decryption should be infeasible without the private key.  

Both secret-key and public-key encryption schemes generally include some randomness when encrypting, so that different encryptions of the same message are not the same.

\emph{Secret sharing}~\cite{shamir1979share} allows a secret such as a key to be shared among several participants so that it can be recovered only if some threshold meet and exchange their shares.  Fewer than a threshold of participants can derive no information about the secret.

\subsection{How cryptographic hashes of names can be reversed}\label{sec:hashing}
There is a persistent misunderstanding in the PPRL literature that cryptographic hash functions are impossible to reverse. This is incorrect. 
Irreversibility can be true only if the input is \emph{randomly and uniformly} chosen from a sufficiently large set that it is infeasible to try them all to see which one matches the given output. 
Names are clearly not chosen in this way.  The next sections explain some known ways of reversing hashes when the input set is predictable, as names are.

\subsubsection{Dictionary attacks}
Name reversal can be applied to an entire database given a list of all (or many) probable names, derived for example from the Whitepages or the 2021 Census, or simply from the attacker's memory of known names.

Simply trying all possible inputs, as described in Section~\ref{DictionaryAttack} is known as a \emph{dictionary attack}. Modern security would require at least $2^{128}$ possible input values to be considered secure against a brute force (dictionary) attack. There are fewer than 400,000 last names currently in use in Australia, which is small enough to guess all possible values.  Calculating cryptographic hash values for all of them would take mere seconds.  We ran a  demonstration of this in our seminar at the ABS, based on simply running through a directory of all Australian names\footnote{We used the surname list from IP Australia and a list of baby names.} to see which one matched a SHA-256 hash of a volunteer's name.  It took a few seconds to recover.   Cryptographic hashing alone provides a near-zero level of security in this context.

\subsubsection{Why plain HMACs do not solve the problem} 
More recent proposals have replaced the plain cryptographic hash with an HMAC (Hashing-based Message Authentication Code). A number of papers incorrectly refer to this as a hash, when it is not. It uses a hash but has different properties and security guarantees. The most significant difference is that it requires a key $k$. The key must be generated in accordance with cryptographic procedures with good entropy.

The key is critical to the security of the output and must be kept absolutely secret. Were someone to gain access to the key, the HMACs tags (output values) would become as easy to reverse as plain cryptographic hash. A number of papers reject  encryption as inappropriate because it permits decryption with knowledge of the key. Yet the same is true if using an HMAC over a small input set (such as names), for the reasons described above.

The misunderstanding of HMACs gives the mistaken impression that they comply with the requirement to be irreversible, when actually they do not.
In most cases, the input sets are small and the security of HMACs reduces to being equivalent to encryption. There is certainly no guarantee of irreversibility---just as with encryption the security depends on the key and access to the key. 

\subsubsection{Determinism and frequency attacks}\label{sec:freq}
Another vulnerability of HMACs is that, for a given key, all HMACs of the same message are equal.  (Note that this is not generally true of encryption, though it is true of keyless hashing too.) Such approaches allow efficient exact matching via comparison of outputs. However, whenever a deterministic approach is used, the frequency distribution of the input is replicated in the output. This presents a particular problem where the input distribution is not uniform, as is the case with names. For example, the name ``Smith" is overwhelmingly the most popular last name in Australia. By looking at the output HMAC tags it would be trivially easy to identify which one represented ``Smith" \emph{even without knowing the secret key}. In some cases, as in \cite{ONSM9}, where similarity information is provided, being able to reverse a single encoding can lead to reversing of many more \cite{culnane2017vulnerabilities}. 

Even where schemes apply a further level of abstraction, as is the case with Bloom filters, it has been shown that frequency analysis can still be performed and  used to recover plaintexts \cite{kuzu2011constraint}. 

So plain HMACs can be reversed if the attacker either
\begin{itemize}
    \item knows the key and can guess the input value (for example by iterating over all possible input values), or
    \item doesn't know the key but can identify an input by the matching frequency of its output.
\end{itemize}

\newpage
\section{ABS requirements}\label{sec:requirements}
Much of the Privacy-preserving record linkage literature is concerned with ``... how two or more organisations can determine if their databases contain records that refer to the same real-world entities without revealing any information besides the matched records to each other or to any other organisation.''\cite{christen2012data}  ABS's setting is slightly different, because  ABS is a single party aiming to link disparate datasets within the organisation, under the assumption that it arrives at the ABS in plaintext, but is then encoded to limit the 
 recovery of the underlying name from the encoded name.  Thus the ABS requirements are different from the usual requirements in the literature.

The following is a summary of the key requirements as captured on 7th December 2016 during discussions between ABS and the University of Melbourne.  Functional requirements capture what the protocol ought to allow properly-authenticated ABS employees to do; security requirements capture what the protocol ought to prevent an attacker from doing.

\subsection{Functional requirements}
\begin{description}
    \item[Link First and Last Name] The approach should provide a way of linking the first and last name fields. The first and last name should be treated separately. Address will be handled via geo-coding. 
    \item[Fuzzy Matching] Ideally the approach would provide for inexact matching to handle typical data capture errors such as transposed characters and differences in spelling.  Note: this is a desirable but not necessary requirement, since names could be canonicalised before matching.  
    \item[Exact Matching] The matching should aim for an exact match. This is on a data level, as opposed to a record level i.e. Bob Smith matches to Bob Smith, but there may be multiple records with the name Bob Smith.  Note: this again is a desirable but not necessary requirement, since a one-to-many matching of names could still allow one-to-one matching of records given other information.
    \item[Integrate into Data Integration Protocol] Ideally the approach will fit with the Data Integration Protocol. Whilst the protocol is not absolutely rigid, and could be modified, any modification would require an equivalent business case.  Cryptography could be used to enhance security of the data integration protocol by enforcing existing rules that restrict the data visible to different participants.
\end{description}

\subsection{Security requirements}
A key part of this project is translating into mathematical terms the requirements for security and privacy of the linking process.

\begin{description}
    \item[Deletion of Names and Addresses] The ABS has committed to the deletion of the names and addresses. The Senate submission includes the statement that ``ABS confirms names and addresses will be destroyed when there is no longer any community benefit to their retention or four years after collection, whichever is soonest''
    \item[Cryptographic Hashing] The ABS has made a public commitment to using a Cryptographic Hashing function. The statement reads ``ABS will use a cryptographic hash function to anonymise name information prior to use in data linkage projects. This function converts a name into an unrecognizable value in a way that is not reversible...''.
\end{description}

Taken together and in an absolute sense, these requirements are impossible to deliver. 
The requirement to delete names and addresses, if taken in an absolute sense, would include deleting derivatives of the names, which would prevent linking. It would be clearer to distinguish between plaintext and encoded names, and only assert the deletion of the plaintext names. 
The assertion that a cryptographic hash cannot be reversed is mathematically incorrect
in this setting, as explained in Section~\ref{sec:hashing}.

For complex unit-record level data, including Census data, auxiliary data can be often be used to link individual records, even without name.  This is the basis of ABS’s existing bronze linking.  
This means that they can probably be re-identified without the encoded name anyway.  Protection against re-identification depends on good processes within ABS.

The undertaking on the encoding of names should therefore be considered in the full context of auxiliary data and ABS processes.  There are several reasonable interpretations:
\begin{enumerate}
	\item	That the encoding cannot be reversed except with a secret key held by ABS.  This is the property achieved by encryption (Option 1), if properly implemented;
\item	That the encoding, taken alone without auxiliary data, cannot be reversed to a unique name.  This is the property achieved by lossy encoding (Option 2), if properly implemented;
\item	That the encoding doesn't make re-identification easier, or increase the number of records that can be re-identified, except with a secret key held by ABS.  This is the property achieved by  HMAC-based linkage key derivation using subsets of attributes (Option 3), if properly implemented.
\end{enumerate}

\vjt{``Shared'' sounds like the opposite of what it should mean.}
Using encryption, for example, would mean that ``not reversible'' must be reinterpreted to mean ``not reversible except given certain secret keys.''  These keys would need to be stored securely or secret-shared among several entities within ABS---much like is already done in the ABS Data Integration Protocol.  The security of such an approach is based on the assumption of trust and compliance with a process or protocol for key management.   The distribution of trust could possibly be designed to align with existing protocols---this is the topic of the next sections of this report.

The following sections describe the tradeoffs among the various options for linking.  For each option, we discuss how it addresses both the security requirements and the functionality and efficiency requirements.  Some of the options could be combined.  For example, if encoded names needed to be stored for longer periods, they could be generated using HMACs on subsets of attributes (Section~\ref{hmac_linking_keys}) but then encrypted with the public key of the Linker for storage (Option~\ref{sec:encrypt}).

\section{Option 1: Encrypting names using public-key encryption} \label{sec:encrypt}
The simplest secure approach is to encrypt names with the public key of the Linker.\footnote{In practice, public key encryption can be implemented by generating a random key for a secret-key algorithm such as AES, encrypting the data with that key, and then encrypting the key with the recipient's public key.}  Other linkage items such as year of birth and location could also be encrypted, which would improve the security of the whole system.  This would scarcely alter the ABS's existing linkage process at all, except that the Linkage File produced by the Librarian would be encrypted.  Indeed, the whole process could be considerably simplified: rather than a separate manager for names, there could be an initial step in which the names and any data used for linking were encrypted.  The data could then be stored in encrypted form and simply passed to the Librarian and on to the Linker whenever linking was performed.

Variables that are both linking variables and analysis variables (such as year of birth) could either be sent separately to the Assembler (as they are now), or decrypted by the Linker and sent to the Assembler with the Linkage Output File.

\begin{description}
\item[Information required to make the anonymised name/linkage file: ] the public key of the Linker; the names and (possibly) other linking variables.
\item[Information required for linking:] the private key of the Linker. 
\item[Information required for reversing:] the private key of the Linker.
\item[Ways of inhibiting unauthorised reversing or linking:] keep the Linker's private key secret.  For example, it could be secret-shared among several people at ABS or even some people outside, so that many had to participate to decrypt.   Depending on how the SBTP access control mechanisms are implemented, it may be possible to simply re-use their key management infrastructure.
\item[Fuzzy matching: ] Yes, by the Linker, after decrypting the data.  Any fuzzy matching algorithm could be applied.
\item[Linking accuracy: ] As good as linking on unencrypted values, because the Linker can see all unencrypted values.
\item[Implementation difficulty: ] The protocol for encrypting, decrypting, and managing keys would need to be implemented with care by professionals, but would use standard techniques and might be able to re-use methods from the SIAM/SBTP technologies.
\item[Computational Efficiency: ] This needs to be tested, but would probably be very efficient in practice because the cryptography is simple encryption/decryption, and the linking is performed on unencrypted values.
\item[Other advantages: ] In future, other agencies sending their data to ABS could also encrypt their names and data with the Linker's public key.  This would mean that nobody within ABS would see the incoming names, except with access to the Linker's private key.  It would also protect the names and data from compromise during transmission between agencies.  Although the private key must be kept secret, it would not have to be distributed very widely ever.  In particular, it is not needed for the \emph{production} of the encrypted files---that is the great advantage of public key cryptography.
\item[Other disadvantages: ] Strong requirements for keeping the private (decryption) key secret.
\end{description}

\section{Option 2: Lossy encoding for names} \label{sec:lossy}
Lossy encoding creates a many-to-one mapping between inputs (names) and outputs (encoded names). 
The idea is to encode first and last names, separately, using a bucket that includes a whole set of possible names.  A simple example is retaining initials rather than whole names.  This is impossible to reverse without auxiliary data because many names have the same initial\footnote{Though if only one person has a particular pair of initials, more information than this needs to be lost.}.  It nevertheless provides some useful information for linking, because a particular set of initials won't match most names.  All methods for lossy encoding have this same logical structure: information is truly lost, but some information is retained which can be used to eliminate some incorrect matches.  Of course, the information that is retained also helps an attacker do unauthorised linking/re-identification if the dataset is leaked.

This option leverages the redundancy of identifiable information---most people are unique based on a subset of the attributes usually associated with Census data, so deleting some information from the name may not severely reduce the accuracy of linkage.  The Linker would also need to consider age, gender and country of origin or last residence for accurate linking.
Unfortunately, this feature is also a challenge for privacy---attributes that can be used by the Linker for more accurate linking could also be used by an attacker for unauthorised re-identification.  This approach therefore relies on proper processes within ABS for keeping the dataset secret.  

Different lossy encodings might retain very different amounts of information.  For example, the first three letters of a name provide much more information than the initial alone.  The amount of information lost affects (legitimate) linking quality as well as the likely success of an attacker.  

One approach to lossy encoding is to define a function that maps input names to output buckets, for example using the
ASCII character codes. Ideally, the function should have a near uniform output. One way to achieve this would be to use an HMAC and truncate it to an appropriate length for the number of buckets. For example, if you wanted 256 buckets you could truncate to the first 8 bits, and treat it as an integer. 

The downside of the simple approach above is that it will to some degree replicate the frequency distribution of the input in the output, particularly for high frequency values. For example, the bucket with ``Smith" will be detectable by looking for the most popular bucket (though other names may map to that bucket too).  We simulated this approach using between 10, 50, 100 and 500 buckets.  Smith was in the largest bucket every time.

One approach to mitigating this is to apply a more structured mapping that attempts to smooth the frequency distribution of the output. Such a mapping could combine less popular names together to create output buckets that are broadly of equal size. In such an approach the Name Manager produces, and the Librarian and Linker use, a table that lists the correspondence between name and bucket. Without that list, it should be hard to know what bucket contains what names.  

One step of privacy protection is to keep the name-bucket correspondence secret; the other step is that many names have the same code, so extra information (such as age or location) is required to link an individual record.

However, this protection is quite easily reversed given some information about some people.  Once one person has been re-identified based on other attributes such as their location and date of birth, the adversary can infer that everyone else with the same name is in the same bucket.  It would only take a little auxiliary data, with a few successful re-identifications, to learn at least part of the name-bucket correspondence.   

\begin{description}
\item[Information required to make the anonymised name/linkage file: ] the names.
\item[Information required for linking:] The table or function associating each name with its bucket, and also some other linkage variables such as age or location.
\item[Information required for reversing:] The name-bucket table, plus also some auxiliary information about the other linkage variables.   
\item[Ways of inhibiting unauthorised reversing or linking:] Keeping the name-bucket table secret; encrypting other linkage variables.  Unfortunately, unauthorised reversing or linking would be straightforward unless the other attributes were all encrypted.
\item[Fuzzy matching: ] Possibly, depending on how the buckets were assigned.  Fuzzy matching would come automatically if similar names could be assigned to the same bucket.  However, it would be very difficult for names that were similar but assigned to different buckets.
\item[Linking accuracy: ] Less than for full name matching.  This could see an increase in false positives, particularly as a result of any overt frequency smoothing that has been applied.  Accuracy would depend on how many other linkage variables were available. 
\item[Implementation difficulty: ] Depends on the method of generating the name-bucket table.  This table would need to be stored in a secure way. The functional approach is simpler but remains susceptible to frequency attacks.
\item[Computational Efficiency: ] Generating the table is probably quite efficient, depending on the method of generating the name-bucket table.   However, the efficiency of the linking process could suffer because of the increased rate of false-positive matches.
\item[Other advantages: ] Compliance with a literal interpretation of a name encoding that ``cannot be reversed'' to a unique name, if properly implemented.  
\item[Other disadvantages: ] Reduced accuracy of linking.
\end{description}

\subsection{An analysis of name frequencies and the implications of incorporating other variables} \label{sec:freqsmooth}

As discussed above, most lossy encoding techniques remain susceptible to frequency attacks unless they are deliberately designed to produce buckets of equal size. In this section we look at how that could be achieved, and discuss some of the issues associated with it. 

\subsubsection{Input distribution equals output distribution}
\begin{figure}[htb]
\centering
\includegraphics[width=\textwidth]{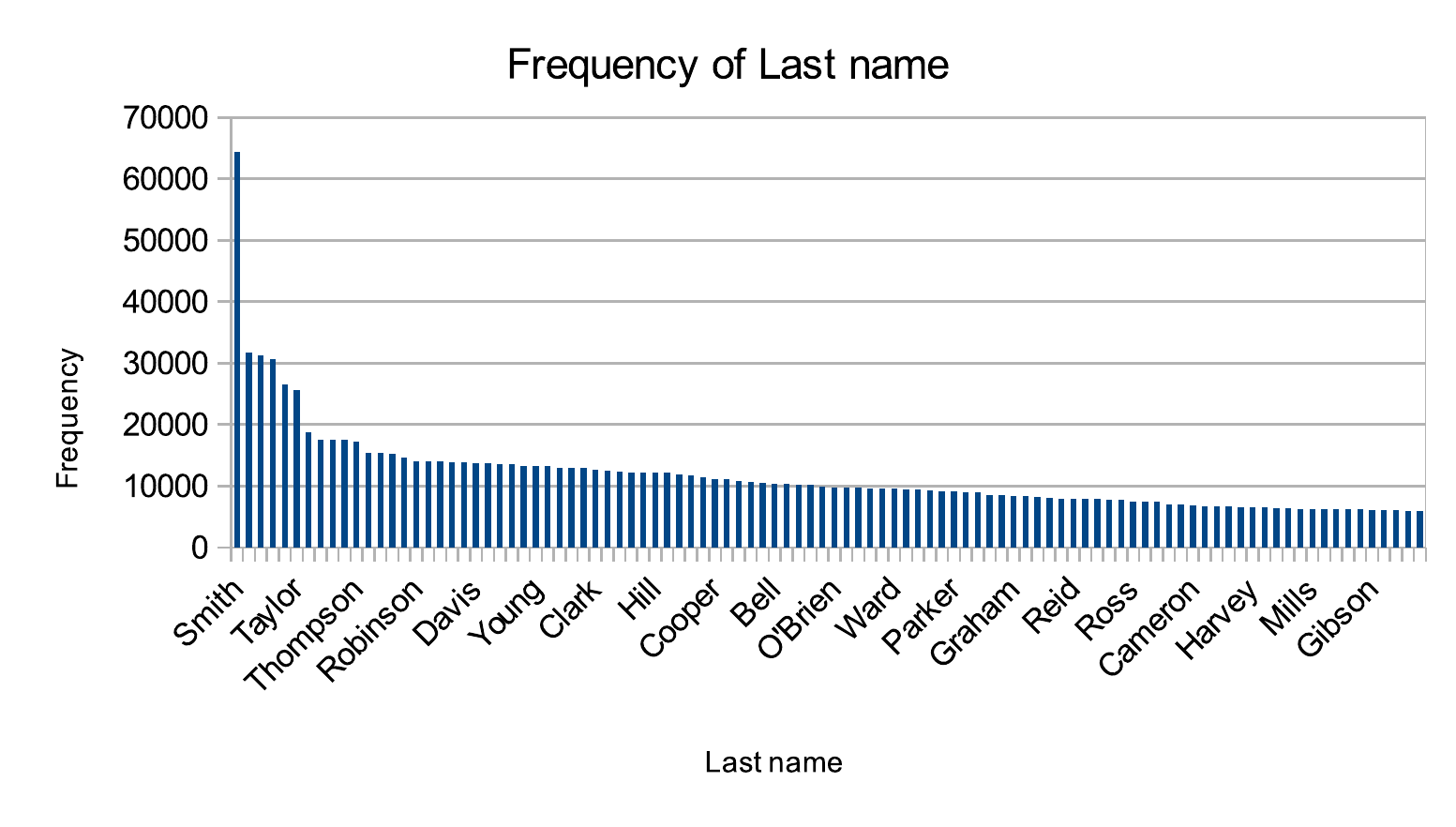}
\caption{Frequency of last name}
\label{fig:freq_last_name}
\end{figure}

If the mapping is deterministic, in that the same input is mapped to the same output each time, the input frequency distribution is largely replicated in the output. This is exactly what we want to avoid, since it leaks information and risks breaking down many-to-one relationship, at least probabilistically. For example, ``Smith" is overwhelmingly the most popular last name, and whatever bin it is assigned to will be proportionally more popular than any others. As such, any row assigned to that bin has a high probability of being ``Smith''. Attempting to create a uniform output distribution will likely lead to significant loss in accuracy. The frequency of ``Smith" cannot be subdivided into multiple bins without increasing the false negative rate, so the only way to smooth the output distribution is to combine output bins to create the same large frequency. Looking at figure \ref{fig:freq_last_name}, which shows the frequency distribution of last names in Australia it is clear that the dominance of ``Smith" is an issue. In order to achieve a uniform output, or something close to uniform, it will require combining many low frequency names into a single bin. Even  combining high values together, for example, ``Jones" and ``Williams" (the next two most popular names) will still result in a bin containing nearly 64\,000 unique names. This will give high accuracy to popular names, but very poor accuracy to everything else. The skew in the input distribution is just too significant to meaningfully smooth it without a significant loss of accuracy.

\subsection{Summary}
The main advantage of lossy encoding is a literal adherence to the promise that the encoding of names ``cannot be reversed.''  Lossy encoding itself doesn't significantly mitigate the risk of unauthorised linking---the same auxiliary information used for ABS linking could be used by an unauthorised attacker.  It is therefore still very important to encrypt or otherwise protect the other  variables.

\section{Option 3: HMAC-based anonymised linkage identifiers  using subsets of attributes}\label{hmac_linking_keys}

In this section we describe a method for combining multiple attributes into a single anonymised linking identifier.  This has many advantages for both computational efficiency and privacy.  There is some degradation in linking quality compared to Option~1, but this may not be significant depending on how the linking identifiers are chosen.  It could also be combined with a lossy encoding of names if required.  The main advantage of this approach over plain lossy encoding is that the linking could be performed on the anonymised linking identifiers by a Linker that didn't need to know the decryption key (though this would require some modifications to the current process).

\subsection{Smoothing the input distribution by including multiple attributes}
We want a distribution in which all inputs are unique. We can then assign those values to different outputs to maintain both privacy and accuracy. The only way to achieve this is to include many attributes in the input value. Combining first and last name will have some effect, although it will still display a skew. For example, there will be more  ``Steve Smith''s" than ``Shanika Karunasekera''s.  Also, birthdate correlates with first name because  first names follow fashions that change over time. 
The best idea is to combine many more fields into the input and then create multiple anonymised linking identifiers in an approach similar to that used by the UK Office of National Statistics (ONS). In \cite{ONSM9}
they create 11 anonymised linkage identifiers with various different parts of first, middle and last name, date of birth, postcode, and gender. They report a uniqueness value of at least 98\% for all the linkage identifiers. The ONS do not perform a lossy encoding on those attributes, instead matching on them directly, but they could be lossy encoded.  

\subsection{The cryptographic construction}
An HMAC is a function that takes a message and a secret key and returns a digest (often called a hash).  We have explained elsewhere that, even without knowing the key, the HMAC of a list of names can be reversed because of frequency attacks.

Using unique inputs is a good way of mitigating frequency attacks on name-based HMAC. If you incorporate enough extra data, every record should be unique. Without the key, it could not be reversed (because an HMAC behaves like a random function in this case~\cite{bellare2006new}). With the key, it could be reversed only by knowing (or guessing) all the attributes in one hash/encryption.

The idea is similar to a technique in use by the UK ONS \cite{ONSM9}. It requires the secret key to be securely generated and carefully protected. The idea is, for each record, to produce several encodings using different combinations of variables (combined using the secret key) and store them all. For example, one might use first name, DoB and address; another might use surname, DoB and country of last residence. 

For example, if $k$ is the secret key then the Linkage File for a particular Link ID could be computed as
$$ \begin{array}{rl}
\textit{Digest}_1     &  = \HMAC_k(\textit{first-name}, \textit{address}, \textit{birth-year}) \\
\textit{Digest}_2     &  = \HMAC_k(\textit{first-name}, \textit{last-name}, \textit{address}) \\
\textit{Digest}_3     &  = \HMAC_k(\textit{country-of-last-residence}, \textit{address}, \textit{birth-year}) \\
\textit{Digest}_4     &  = \HMAC_k(\textit{first-name}, \textit{last-name}, \textit{birth-year}) \\
\end{array}
$$
or whatever other combinations of variables seemed useful.  

When a new database is linked, the same computation is repeated on the incoming variables.  
If the person has changed address, for example, the digests that use address will not match, but other digests should.  If their surname has been mistyped, then the digests that don't use surname or only use the first two letters might match.
We assume the Linker has access to the plaintext of the dataset to be linked.  So information about how common each name is, or how likely it is that a certain name has been mistyped, {\it etc.} could be derived from the non-Census data. Then it makes a linking decision based on how many collections of attributes seem to match and which ones it expects to have changed.

This technique could be combined with preprocessing of names for fuzzy matching, for example the n-gram approach of \cite{schnell2009privacy}, or with known common transcription errors such as the reversal of names.  The linker could try likely misspellings of names at linking time if the given one didn't match.  For example, given an input name ``Smithe'' the linking could be attempted using ``Smithe'' and, if it failed, re-attempted with ``Smith'' then ``Smythe'' {\it etc.}    The same technique could be applied to other variables that may not quite match.  For example, when comparing 2021 ages with 2016 ages, you could subtract 4,5 and 6 years from the ages before recomputing the hash/encryption. Dealing with typographical and transcription errors in names is harder, because you need to guess what they were. 
Name standardisation should help but may never produce results as good as having both names in the clear.

Ideally this would be handled by careful selection and construction of the encodings. The attributes should be selected to handle the typical distortions seen in the datasets. Where the above would potentially be useful is where there are compounded distortions, {\it e.g.} someone has moved, changed their name and mistyped their firstname.

Obviously the anonymised linkage identifiers are not conditionally independent---if one attribute changes, then several linkage identifiers  might change.  This complicates the analysis for matching---it means that when a record matches some, but not all, linkage identifiers, a careful inference must be made about which attribute(s) might have changed.  In Deterministic linking, multiple potential links will not be linked. In probabilistic linking, quality measures usually depend on the strength of each linking variable. This needs to be adapted to give a quality measure to each collection of variables.  Also, mismatches among independent collections should be regarded as much more important than mismatches among related collections.  For example, if every identifier involving address fails to match, but all the other ones do match, then the person has probably moved; if a similar number of mismatches occur, but do not all have a common variable, then at least two variables must have changed and it is less likely to be the same person.

\begin{description}
\item[Information required to make the encoded name (and other data) file: ] The HMAC secret key, the names and other linkage variables.  (Note that this technique only works by combining names with at least some other variables.)
\item[Information required for linking:] The HMAC secret key plus some name and linkage variables from the incoming dataset.
\item[Information required for reversing:] Either frequency information on collections of variables (we would aim for this to be nearly uniform) or the HMAC secret key combined with a successful guess of at least one collection of variables. 
\item[Ways of inhibiting unauthorised reversing or linking:] Keeping the key secret; ensuring that the encodings incorporate several variables.
\item[Fuzzy matching: ] Yes, the linking identifiers already provide some fuzzy matching.  The ONS report a very high rate of matching on the linking identifiers alone. 
Furthermore, if a perfect match was not possible, the incoming names could be slightly perturbed and retried.  This would not be quite as accurate as plaintext name matching, but could be quite good in practice. 
\item[Linking accuracy: ] It depends on which attributes are included in the HMAC digests.  This would need some empirical investigation, the greater the degree of uniqueness the better the accuracy. A key which does not provide sufficient uniqueness not only risks privacy through frequency attacks, but also impacts on accuracy by causing false positives. 
\item[Implementation difficulty: ] Similar to encryption.  It would need careful generation and management of the HMAC key and professional implementation of the cryptography.  Possibly it could use existing libraries from the SIAM/SBTP project---this would need to be checked.
\item[Computational Efficiency: ] Currently the most efficient approach we have (more efficient than plaintext similarity matching). This is largely due to it being deterministic matching, allowing extremely efficient matching on very large sets without requiring cross comparisons. It is efficient enough that it may not require any blocking, allowing whole population linking. 
We will discuss further in a later section. We require further analysis to establish the accuracy of the approach on very large sets.
\item[Other advantages: ]  The main advantage of this over simply applying HMAC to each name separately is the mitigation of frequency attacks, if properly implemented.  It also means that, even if the HMAC secret key was compromised, an attacker would need to guess all of the attributes for one of the digests in order to recover the name.
\item[Other disadvantages: ] This structure would not be directly useful for computing the statistical data necessary for assessing the accuracy of probabilistic linking.  Some of those values could be computed independently, but the ones involving names could not.
\end{description}

\subsection{Defending against frequency attacks}
Even for an attacker without the key, HMACs are subject to frequency attacks.  The technique described in this section is secure only if a large enough collection of attributes is chosen to make the linkage identifiers entirely unique.  Section~\ref{sec:freq} described frequency attacks as applying to very frequent names, but the same problem occurs in sets of almost-unique identifiers with one or two repeated values.  Suppose for example that two people with the same first and last name live at the same address, which happens occasionally in cultures where children are named precisely after their parents.  Then a digest incorporating first name, last name and address would be almost entirely unique except for those households.  This would allow those individuals to be isolated among a small set of possible records.

At the time of building the digests, it is critically important to check for duplicates.  If there are any, then records should be removed until all the digests are unique---this obviously lowers linking accuracy, but is critical for preventing frequency attacks.  If there are too many duplicates to remove, then that collection of variables should not be used for a linking identifier.

This is the reason that a fairly large number of attributes need to be included in each digest.  Empirical testing, on the spot, could determine which collections produced unique (or close enough to unique) outputs.

\subsection{Deterministic linking - performance advantages}
 The linking identifiers could be stored in a database, with the corresponding original recordId as a field. Since most of the linking identifiers are unique they constitute an excellent record identifier, which can be effectively indexed. This is a critical performance advantage in practice: a database index allows the record to be found in a single lookup, rather than by searching through the entire list of millions of values until a match is found.  When performing the linking, we only need to iterate over the incoming records and perform a single query for each of its record linking identifiers. Those queries are extremely quick because they are just looking up an index.  
 
 The database could be used directly as both the Anonymised Name File and the Linkage File in ABS's current process.  It also obviates the need for a separate Linkage Concordance file, though this could be included if desired.

One of the advantages of this approach is that it allows deterministic linking, whilst still handling some degree of distortion. Deterministic linking is considerably more efficient because it can be achieved by indexing the database by each encoding, thus avoiding a full cross comparison of the two datasets.

By way of an example, if we wanted to link the 2.9 million records in our sample dataset it would require a maximum of approximately 32 million queries. On a multi-core desktop machine we are able to perform those queries and the necessary linking in under 15 minutes. If we compare that to any scheme that involves cross comparison, we would have to perform $(2.9 \text{ million}) ^2 / 2=4205000000000$ record comparisons. Even if we could perform each comparison in a microsecond, on an 8 core machine, it would still take over 6 days. In such circumstances, blocking is essential to allow the linking to be feasibly performed.

However, blocking has its downsides, namely, that it will impact on accuracy. For example, if a geographically based blocking algorithm is used, and an individual changes address to somewhere outside of their block, they will definitely not be matched. This can be mitigated somewhat by performing multiple passes of blocking on different attributes, but it will still have an impact. \DE{We do have multiple passes to relax geography so that we can pick up the people who moved or are travelling. Blocking is not a limitation so long as the blocking and linking strategy is designed well for the data quality and the objectives} 

\subsubsection{Whole population linking}
When looking at the figures above, it becomes apparent that the deterministic linking identifiers approach is efficient  enough to perform whole population linking. This would both simplify the implementation and also avoid the negative impacts of blocking. However, it will be essential to ensure the linking identifiers remain unique across the entire population, and importantly, adequately handle the expected distortions. 

Ideally we could get some data to evaluate the rates of uniqueness (which should be very high) for combinations of first name, last name, DoB, address, country of origin/last residence. Then also investigate which subsets of attributes are also (almost) all unique.

This creates the rather unintuitive situation of providing better privacy by adding more information about someone. An important caveat is that any dataset that is to be linked must provide the same granularity of data in order to create the necessary linking key. For example, if year of birth is included in the original dataset, but the incoming dataset doesn't include that attribute, then none of the linking identifiers that incorporate year of birth can be used.  Other linking identifiers that do not contain year of birth can. The decision of what attributes to use is vital and would need to be driven by the data that is available and the level of uniqueness it offers.   
If it is necessary to perform the same pre-processing as the ONS approach, it would make more sense to utilise their approach for linking, instead of apply a further lossy encoding that may not improve privacy much, but could impact on accuracy. 

\section{Option 4: Individual IDs}\label{individual_ids}
Suppose that each person could be assigned a unique ID number.  Then we could separate out two processes:

\begin{enumerate}
\item the process of linking a particular name, address and date of birth to the ID number, and
\item the process of linking the records associated with that ID number across different databases.
\end{enumerate} 

So suppose that the ABS (and other agencies) had a large table like this:

\begin{tabular}{llll}
{\bf Name} & {\bf Address} & {\bf DoB} & {\bf ID num} \\
John Citizen  & 1 Tree St Broadmeadows & 10 Jan 1970   & 5795935 \\
Jane Citizen    & 5 Apple Rd Surrey Hills & 25 Dec 1912 & 12334225 \\
... & ... & ... & ...\\
\end{tabular}

This table is not intended to be secret or sensitive: it is like the whitepages, with a link to a non-secret ID like a tax file number or the US social security number.

The suggestion in this section is that stored datasets remove the names, addresses and dates of birth entirely, and store instead the ID number, encrypted so that the private key is secret-shared among multiple people.  When a new dataset is received, the ABS should first link the name, address, and date of birth to an ID number---this uses no cryptography, just whatever techniques for fuzzy matching ABS is familiar with.  When each record has been assigned to an ID number, the names, address and dates of birth should be removed---the rest of the linking process should occur by looking for exact matches of the ID number.  This can be performed on encrypted values.

The assignment and encryption of ID numbers could also be done by other agencies before they send data to the ABS.

\subsection{Methods for linking records based on exact ID matches}
Camenisch and Lehmann \cite{camenisch2015linkable} describe a protocol for linking individual ID numbers across government databases, in a very strong security model in which 
three different parties cooperate to perform linking while leaking very little information about individual identities.  This would be a good starting point for a design of a protocol for future use. Their setting is:

\begin{itemize}
    \item each of several data authorities may have a public key and some data,
    \item each person has an ID number,
    \item a linking authority knows some master information that allows it to translate IDs encrypted for one data authority into the same ID encrypted for a different data authority.
\end{itemize}

Linking is performed by performing blinded decryption - a process in which a random shift is first applied to the cipher so that the decryption is the real value plus an unknown blinding factor. This is effective for exact matching, since if the two plaintexts are equal and you apply the same blinding factor to both, they will still be equal when decrypted. If they are not equal nobody learns what the original ID was. 

This protocol has many good properties.  In particular, it is ``blind'' in the sense that the linking authority does not learn which ID it is linking.  

\begin{description}
\item[Information required to make the anonymised name/linkage file: ] The table linking IDs to the name, address, {\it etc.}, The public key of the Linker.
\item[Information required for linking:] The Linker's private key, which can be secret-shared.
\item[Information required for reversing:]  The Linker's private key and the ID table, which is public. 
\item[Ways of inhibiting unauthorised reversing or linking:] Keep the Linker's private key secure
\item[Fuzzy matching: ] Yes, at the stage where a name/address/DoB is matched to an ID.
\item[Linking accuracy: ] Could be very high, becase fuzzy matching is performed on cleartext names.  The fuzzy matching would, however, be in two steps, effectively canonicalising a name each time.
\item[Implementation difficulty: ] Complex and requiring careful cryptography.
\item[Computational Efficiency: ] Feasible but taking longer than plain encryption.
\item[Other advantages: ]  
\item[Other disadvantages: ] 
\end{description}

\section{Option 5: Homomorphic encryption} \label{hom}
\emph{Homomorphic encryption} allows certain computations to be performed on encrypted data.  For example, it has been used to add encrypted votes and then decrypt only the totals, not the individual ballots~\cite{adida2009electing}.  Recent advances in cryptography include more efficient algorithms for wider classes of computation.

In principle it is now possible to compute any function (for Linking, comparison, {\it etc.}) on encrypted data, decrypting only the final answer.  In practice, however, the most general techniques require an impractical amount of computation.  It is not practically possible to compute a rich linking process, tolerating fuzzy matching and other issues, in a reasonable time.

However, it would be possible to implement some simple comparisons on encrypted names, such as computing the Hamming distance (the total number of different characters).   In this case names could remain encrypted throughout the linking process, while only distances were decrypted.  The keys used to decrypt the distances \emph{could} still be used to decrypt the names, but in a proper linking process they never would be.  

This process would be very secure because the decryption key would never need to leave the Linker.   It could even be shared among multiple people so that encrypted names truly could not be reversed unless a threshold of those shares were compromised.

However, even this restricted notion of homomorphic encryption would require vast computational resources.  This, combined with the restricted set of linking policies, make it an unattractive option at present.  However, it may be worth revisiting in the future if techniques improve.  It could be combined with a secure multiparty computation (SMC)  technique for computing only the links without revealing the inputs (See Literature Review Section~\ref{sec:lit:smc}).  Indeed, many  MPC protocols use homomorphic encryption.

\begin{description}
\item[Information required to make the anonymised name/linkage file: ] The public key of the Linker.
\item[Information required for linking:] The private key of the Linker, but this could be stored in a distributed way and never explicitly recombined.
\item[Information required for reversing:] The private key of the Linker.
\item[Ways of inhibiting unauthorised reversing or linking:] Keeping the Linker's private key secure; never explicitly computing it.
\item[Fuzzy matching: ] Provides a modified hamming distance which will provide fuzzy matching equivalent to using such a string comparison metric. Could be improved further by careful string encoding.
\item[Linking accuracy: ] Multiple comparison can be run, i.e. transposing first and last name if they don't match. Accuracy likely to be high and close to plaintext matching.
\item[Implementation difficulty: ] Very complex.
\item[Computational Efficiency: ] Requires intensive computation.
\item[Other advantages: ]  This is a very secure option, because the decryption key would never need to leave the Linker.  It could even be shared among multiple people so that encrypted names truly could not be reversed unless a threshold of those shares were compromised.
\item[Other disadvantages: ] Would need careful analysis of what information could be obtained by the Linker from multiple runs of the protocol and measuring the similarity between different names. 
\end{description}

Schemes based on homomorphic encryption or secure computation are the future of secure data processing.  These sorts of schemes are an active area of cryptography research with many applications.  These sorts of schemes would allow ABS to say truly that it was not able to reverse data if the key could be shared among other organisations.  In the long run, these sorts of approaches should become the norm.  For now, however, the difficulty of implementation probably means that this is better suited to a longer research project than a practical proposal for this year's census data.

\section{Empirical linking results based on a Synthetic Data Generator}
In order to evaluate the various methods we constructed a synthetic dataset. Our aim was to create something that mirrored, as closely as possible, the frequency distribution of real world data. Validating this is difficult, since access to real world data is not an option. However, we have based our sampling on real world samples and aggregates.

\subsection{Datasets and frequency distributions}
\begin{description}
\item[MBS Demographics] At the base of the generation is the demographic information from the MBS/PBS release. This contains approximately 2\.9 million records with YOB and Gender.
\item[Last Name] We obtained a list of 384\,370 last names the occur in Australia, and the corresponding frequencies. We draw from this at random with a probability distribution matching the frequencies to append a last name to each row of the MBS demographics.
\item[First Name] We use the NSW data release of frequencies of the top 100 first names for boys and girls from 1952 through to 2015, to select an appropriate Gender and YOB specific first name for each record in MBS demographics. Ideally, we would have more than 100 names, since this only provides 297 distinct boys names, and 377 distinct girls names. Where YOB is not in the NSW release we take the closest year.
\item[Middle Name] We re-use the NSW data, except we draw the middle name from YOB-20. This somewhat arbitrary, but done in an effort to get a different distribution of middle and first names for an particular year. \cjc{I'm keen to expand the first and middle name parts of the generation, but FEBRL doesn't have many more names either}
\item[Mesh Block] We originally used postcode frequency data, but postcode is not fine grained enough to provide an equivalent uniqueness to the UK postcode and therefore achieve equivalence with the ONS results. We subsequently switched to use 2011 Census mesh block population distribution data. We select these at random according the population distribution, providing the mesh block, and a value synonymous with an SA3 area\footnote{We could convert from mesh block ID to actual SA3 codes, but it is not necessary for our analysis, because our distortions are performed only at meshblock level. When using the pseudo-SA3 value we need to just be representative of the number of codes, hence we derive it from the meshblock ID instead of going to the complexity of performing a full look-up} for use in the linking identifiers.
\end{description}

\subsection{Distortions}
We create a number duplicate datasets with distortions applied in order to evaluate the effective of fuzzy matching. The distortion framework is extensible, so we can further or different distortions. We currently apply the distortions to all records in the duplicate and then evaluate overall impact. We have a mechanism for apply these probabilistically as well. The distortions we currently apply are as follows:
\begin{description}
\item[Change Gender] We switch the Gender from M to F and F to M, primarily to simulate typos.
\item[Change Middle Initial] We select a different middle initial at random.
\item[Change YOB] Replace the YOB with a randomly selected YOB drawn from between 1916 and 2016.
\item[First Last Transpose] We transpose the first and last names
\item[Mesh Block Change] We randomly select a mesh block from the same distribution as used in the original generation.
\item[Remove/Add Middle Initial] Remove the middle initial, or randomly add one if there is not one
\item[Transpose Inner Letters of Last Name] We transpose 2 adjacent letters in the last name, picked at random. 
\item[Transpose Inner Letters of First Name] We transpose 2 adjacent letters in the last name, picked at random. 
\end{description}

\subsection{Analysis of HMAC-based anonymised Linking identifiers}
In order to evaluate the effectiveness of the HMAC Linking identifier approach we constructed a dataset of the relevant keys. We subsequently imported that data into a MongoDB database, with one collection per type of identifier, i.e. ForenameSurnameYoBSexSA4 was a collection. Within that collection each generated HMAC Linking identifier was a document, indexed by the HMAC Linking identifier value. Within the document was an array containing the rowID of any record that generated that identifier. The advantage of this approach is that it permits easy indexing of the HMAC Linking identifiers, providing extremely fast look-up times. In most cases the array of matching documents is an array of 1, since the objective is to generate primarily unique linking identifiers. When linking a record, the same set of HMAC Linking identifiers are generated from the dataset to be linked, each one is then submitted as a query to the database to find all the records that match that linking identifier. Such a query takes approximately 5ms to perform. Additionally, there is no cross-comparison, so the number of queries is linear with regards to the number of records being linked. This allows for full population linking to be undertaken on even large dataset. 

\subsubsection{Determining the best match}
The ONS \cite{ONSM9} approach to performing the linking was to use a hierarchy of identifiers, stopping as soon as a unique match was found. Additionally, they removed matches from both sides of the matching. This approach has a number of problems, it weights identifiers resulting in a false positive in one identifier negating all the identifiers below it in the hierarchy. As such, the ordering of the identifiers becomes very important, but difficult to judge. The removal of matches from both sides of the linking also risks compounding errors. In that if matching two equal populations, a false positive will result in either a subsequent false negative (no match found), or further false negative (lower quality match found), since the correct match has already been removed due the first false positive. Additionally, removing records is inefficient in terms of indexing, negating the performance advantages of this approach. As such, we evaluated two different approaches to finding a match.

\paragraph{First unique match}
In this simulation we maintain the hierarchical nature of the identifiers and stop as soon as we get a unique match, i.e. the array of matching rowID's is of size 1. However, in a departure from the ONS \cite{ONSM9} approach we do not subsequently remove the matching record. If no identifiers have a unique match we consider the record to not be matched even, for example, if one identifier matched multiple records. [Note that in a real run we would need to guarantee uniqueness.]

\paragraph{Voting}
The second approach for matching we evaluated was to perform a vote across all identifiers to determine the most likely match. This was calculated by returning the arrays of rowID's and then performing a frequency analysis of the contents. Whichever rowID received the most matches was considered to be a match. If two rowID's had the same frequency one was selected at random. A failure to match would only be returned if there were no matches to any of the HMAC Linking identifiers. This reduces the importance of the order of HMAC Linking identifiers as well as mitigating any identifiers that may have a higher false positive rate, which could be a problem if they appear too high in the hierarchy. 

\subsubsection{Uniqueness}
At the heart of this approach is the concept that the HMAC Linking identifiers are unique.  In order to evaluate that we analysed the identifiers we generated for a synthetic dataset and determined their uniqueness across the dataset. Table~\ref{tab:hmaclinkingkeysunique} contains the uniqueness of the respective HMAC Linking identifiers. AS we can see most of the identifiers provide a very high level of uniqueness, across the  2.9 million records, the lowest being ForenameSurnameYoBSex at 94.524\%. Uniqueness is  essential to privacy protection---any degree of non-uniqueness presents a degree of privacy risk. For example, ForenameSurnameSexMeshblock is 99.998\% unique, however, that leaves a tiny proportion that are not unique. That lack of uniqueness could be caused, in a real-world dataset, by people who are related, for example, a father and son who share the same first name. Such occurrences are rare, and access to auxiliary information could allow an attacker to look for just such rare occurrences. This is analogous to frequency attack, but on a very specific and small scale. One advantage of this approach is that it permits that risk to be quantified and  mitigated. For example, it would be possible to remove entries that are not unique. Such an action would have some impact on recall, but may be preferable to the privacy risk. Such mitigation strategies become difficult when too large a percentage are not unique. For example, consider a linking identifier consisting of just Forename and Surname, which is only 59.778\% unique in our test dataset. Depending on the identifier's location within the hierarchy it could have an impact on precision.  This set of attributes should not be used as a linking identifier.

\begin{table}[htbp]
\begin{center}
\begin{tabular}{|l|r|}
\hline
\textbf{Uniqueness of Linking Keys} & \multicolumn{1}{c|}{\textbf{\% Unique}} \\ \hline
ForenameSurnameYoBSexSA3 & 99.971 \\ \hline
ForenameInitialSurnameInitialYoBSexMeshblock & 99.972 \\ \hline
ForenameSurnameYoBMeshblock & 99.999 \\ \hline
SurnameForenameYoBSexMeshblockTrans & 99.999 \\ \hline
ForenameSurnameYoBSexMeshblock & 99.999 \\ \hline
ForenameSurnameYoBSex & 94.524 \\ \hline
ForenameBiSurnameBiYoBSexMeshblock & 99.844 \\ \hline
ForenameSurnameSexMeshblock & 99.997 \\ \hline
SurnameInitialYoBSexMeshblock & 99.708 \\ \hline
ForenameInitialYoBSexMeshblock & 99.601 \\ \hline
MiddleNameSurnameYoBSexMeshblock & 99.999 \\ \hline
\end{tabular}
\end{center}
\caption{Uniqueness of HMAC Linking Keys}
\label{tab:hmaclinkingkeysunique}
\end{table}

\subsubsection{Matching results}
Table \ref{tab:hmaclinkingkeys} shows the comparison of the matching results for Voting and Non-Voting methods. Precision and Recall are calculated based on comparing the return match with actual match. The linking dataset is a shuffled, and if appropriate, distorted copy of the original dataset. No records inserted or deleted, as such, we would expect recall to be 1. Recall would only drop below 1 when no matching to any record was found. The precision indicates the the HMAC Linking Key approach is an effective method for matching records when faced with the tested distortions. It should be noted that we have not evaluated results based on composite distortions, for example, transposing letters and changing mesh block. However, that would be fairly straightforward to test if required. The robustness of the approach to distortion can be determined by examining the HMAC Linking Keys that are constructed. Effectively, to be robust to a distortion there must remain at least one key that is not impacted by that distortion. For example, where a mesh block changes within an SA3 area we are reliant on the ForenameSurnameYoBSexSA3 and ForenameSurnameYoBSex linking keys to determine matches. Where a mesh block changes outside an SA3area we are reliant on only the ForenameSurnameYoBSex key. This is reflected in the precision results that show mesh block changes cause the greatest reduction in precision. Combining the uniqueness information with expected distortions it is possible to determine whether the linking keys generated will be robust to it, without having to perform an evaluation on the actual dataset. For example, if both Year Of Birth and Gender change we can be certain that no keys will be able to provide a match. 

The set of keys used in our evaluation is not exhaustive, different keys could be created to handle specific distortions, or composite distortions. The only requirement is that they are largely unique. As such, the exact set of linking keys to be used should be derived from the looking at the actual dataset. It is important to perform this step, since once identifying data is deleted additional keys including that data cannot be created. As such, the approach should aim to handle all expected distortions at the point of creation. 

\begin{table}[htbp]
\begin{center}
\begin{tabular}{|l|r|r|r|r|}
\hline
\textbf{} & \multicolumn{ 2}{c|}{\textbf{Non Voting}} & \multicolumn{ 2}{c|}{\textbf{Voting}} \\ \hline
\textbf{Distortion} & \multicolumn{1}{c|}{\textbf{Precision}} & \multicolumn{1}{c|}{\textbf{Recall}} & \multicolumn{1}{c|}{\textbf{Precision}} & \multicolumn{1}{c|}{\textbf{Recall}} \\ \hline
changeInitial & 1.000 & 0.999 & 0.999 & 1.000 \\ \hline
firstLastTranspose & 0.990 & 0.999 & 0.994 & 1.000 \\ \hline
exact & 1.000 & 1.000 & 1.000 & 1.000 \\ \hline
lastName2LetterTranspose & 0.999 & 0.999 & 0.999 & 1.000 \\ \hline
removeAddInitial & 1.000 & 0.999 & 0.999 & 1.000 \\ \hline
firstName2LetterTranspose & 1.000 & 0.999 & 0.999 & 1.000 \\ \hline
meshblockChange & 0.982 & 0.904 & 0.937 & 1.000 \\ \hline
changeGender & 0.987 & 0.999 & 0.967 & 1.000 \\ \hline
\end{tabular}
\end{center}
\caption{HMAC Linking Keys Results}
\label{tab:hmaclinkingkeys}
\end{table}

\subsection{Direct Bi-gram matching}
By way of a comparison we also analysed a simple bi-gram matching process. In this scheme each bi-gram was encoded into an HMAC, with the HMAC then being compared as bi-grams. This provided a degree of privacy protection, although it would remain susceptible to frequency attacks, particularly given the analysis in Section \ref{sec:randvalues_bigramskew} that demonstrated strong skews in the frequency distribution of bi-grams. A major challenge in performing the bi-gram matching was the inefficiency of performing a cross-comparison. It was infeasible to do this for the entire dataset, and as such we had to deploy a blocking procedure. Even with a reasonable level of blocking the processing time was substantial. In order to allow us to evaluate different matching techniques we took a sample of 3 blocks, each consisting of between 15,000 and 18,500 records. We then performed the cross comparison within those blocks. 

We did not evaluate multi-round matching that would involve different blocking methods to allow handling of geographical changes. We only evaluated distortions that would impact on the result, as such, changes to middle initial, age, and gender were not evaluated. Likewise, given that we know that a wholesale geographical change would lead to a precision of 0, we did not evaluate that either. 

In order to determine whether two sets of bi-grams matched we tried two approaches. The simple approach was to calculate a dice-coefficient between the bi-gram sets. This was simple and fast, and maintained the order of the bi-grams. However, it is not robust to insertion or deletions of bi-grams, that wasn't an issue in our tests because we were not performing that distortion, but would be an issue in a real world setting. The second approach was to calculate the q-gram similarity of the two sets of bi-grams \cite{ukkonen1992approximate}. This is calculated by first calculating the q-gram distance, which requires counting the occurrences of each bi-gram in the two strings and taking the sum of the absolute differences of those counts. We then sum the cardinality of the two bi-gram sets and set this as the maximum distance. We then take the calculated distance from the maximum and divide by the maximum to get a similarity score between 0 and 1. 

The disadvantage of this approach is that it is more computationally expensive to calculate, which over a full set of blocks would impact on the time required to perform the linking. 

Table \ref{tab:ngrammatching} shows the results for the first matching approach. We can see that it performs well in exact matching and the letter transposition distortions. It is not quite as good as the HMAC Linking Key approach, but it is not far off.

The transposition of the entire first and last name performs badly in terms of precision, as we would expect, since we evaluate first and last name as distinct values and then combine their similarities. This could be mitigated by performing an additional comparison with the query first and last name transposed, however, this will have the effect of doubling the computational effort required to perform the matching, which could well push a time consuming process into an infeasible process. 

\begin{table}[htbp]
\begin{center}
\begin{tabular}{|l|r|r|}
\hline
\textbf{Bi-gram Linking Dice-Coefficient} & \multicolumn{1}{l|}{} & \multicolumn{1}{l|}{} \\ \hline
\textbf{Distortion} & \multicolumn{1}{c|}{\textbf{Precision}} & \multicolumn{1}{c|}{\textbf{Recall}} \\ \hline
firstName2LetterTranspose & 0.980 & 1 \\ \hline
exact & 0.981 & 1 \\ \hline
firstLastTranspose & 0.001 & 1 \\ \hline
lastName2LetterTranspose & 0.969 & 1 \\ \hline
\end{tabular}
\end{center}
\caption{Bi-Gram Linking Results}
\label{tab:ngrammatching}
\end{table}

The results for the second approach are shown Table \ref{tab:ngrammatchingqgram}. They are marginally worse than for the simpler approach. This is somewhat to be expected, since this matching approach is more tolerant of changes, particularly insertion and deletion. As a result, the chance of false positive increases slightly. 

\begin{table}[htbp]
\begin{center}
\begin{tabular}{|l|r|r|}
\hline
\textbf{Bi-gram Linking q-gram Scoring} & \multicolumn{1}{l|}{} & \multicolumn{1}{l|}{} \\ \hline
\textbf{Distortion} & \multicolumn{1}{c|}{\textbf{Precision}} & \multicolumn{1}{c|}{\textbf{Recall}} \\ \hline
firstName2LetterTranspose & 0.975 & 1 \\ \hline
exact & 0.980 & 1 \\ \hline
firstLastTranspose & 0.001 & 1 \\ \hline
lastName2LetterTranspose & 0.957 & 1 \\ \hline
\end{tabular}
\end{center}
\caption{Bi-Gram Linking Results (q-gram)}
\label{tab:ngrammatchingqgram}
\end{table}

The bi-gram matching approach performs reasonably well, as would be expected. However, the computational cost, combined with it susceptibility to frequency attacks weaken the argument for its use. 

\section{Conclusion}
All good cybersecurity solutions are a tradeoff among different objectives: security, usability, access, accuracy, computation time, cost, {\it etc.}  A clear attacker model is critical for understanding what the security guarantees are, so that the best solution can be chosen.  
The aim of this report is to make clear the assumptions of the protocols, so that ABS's careful processes for managing data security can be accurately matched to the assumptions on which the protocols' security depends.

Detailed unit-record level data, including Census data, can often be re-identified even without the name, based on other information about the person or household, such as birthdates and location.  The promise to encode names using a cryptographic hash function in a way that cannot be reversed is therefore, if taken absolutely, not achievable in the presence of auxiliary data---many records could be re-identified even if the names were completely removed.  

We have presented several options that satisfy reasonable interpretations of the requirements, in the context of auxiliary data and ABS processes for securing Census data.  Option~1 is simple encryption, which (if properly implemented) cannot be reversed except with the decryption key.  Option~2, lossy encoding, sends many different names to the same encoded value.  It can be reversed to a set of names, but not a unique one (if properly implemented), if the attacker has no auxiliary data about the person. Option~3 produces anonymised linking identifiers that do not make re-identification any easier for an attacker who doesn't have the HMAC key (if properly implemented).  It sacrifices some flexibility for very high computational efficiency.  Options~4 and~5 provide suggestions for future directions using some more sophisticated cryptographic approaches based on homomorphic encryption and multiparty computation.


The most computationally efficient solution we could find is Option~3, to compute an HMAC on a collection of different subsets of attributes, then use exact matches at linking time (Section~\ref{hmac_linking_keys}).  This provides some defence against a motivated attacker who does not know the secret key.  The only information leaked is about the frequency of the different inputs --- if the attributes are carefully chosen it is possible to ensure that every input is unique.  
This solution could be adopted, and has approximately the same security, with a lossy encoding of names.

Our literature review explains why some other proposals in the literature, including plain cryptographic hashing and Bloom filters, do not defend against a motivated attacker.

We would like to thank the ABS for their time and engagement in discussing these questions.  We valued the conversations and the motivation to work on a challenging and important practical problem.

A key aspect of earning public trust is to be open about the details of the algorithms used for keeping data secure.  Whichever solution ABS decides to adopt, we hope that this paper  contributes to an open, factual discussion of linking options and census data security.  

\bibliographystyle{plain}
\bibliography{references}

\appendix
\section{Literature Review}\label{sec:litreview}

The field of record linkage has a long history, dating back nearly 50 years~\cite{fellegi1969theory}.  Privacy Preserving Record Linkage (PPRL)  dates back over a decade~\cite{dusserre1994one}. A significant proportion of the literature has been published outside of Computer Science and Information Security venues, with a particular prevalence for publication within the medical domain. 
 This has resulted in proposals not being subjected to the normal level of rigour and analysis associated with information security. 
A number of schemes do not achieve their claimed privacy properties. 
Cryptographic primitives such as hashing have been used without a clear understanding of the security properties they provide and the attacker model they defend against.  Persistent mistakes leading to security problems are repeated in many papers, even when the security issues have been identified. 

Another significant problem is that protocols designed for one attacker model are reused in a different context.  For example, a technique that obfuscates information well enough for well-meaning researchers in a controlled environment may not be sufficiently secure for publishing on the Internet, where malicious and dedicated attackers will attempt to reverse it.

Christen~\cite{christen2012data} provides an overview of various techniques and methods for privacy preserving record linkage. The initial part of Christen's book \cite{christen2012data} covers some of the typical trust assumptions associated with privacy preserving record linkage. Of note is the assumption that when matching data within a single organisation, as in our context, the employees responsible for performing the data matching ``...do not have malicious intents to take identifying or other sensitive information, or the matched data, outside of their organisations for personal gain".  It is crucial, throughout this review of the literature, to distinguish the techniques that  assume that the data will only be available to those without ``malicious intents'' from those that defend against deliberate attack.  Whilst the work covers both hashing approaches and cryptographic approaches, the conclusion is that ``simple one-way hash encoding allows efficient privacy preserving data matching across organisations". 
This is not correct against a motivated attacker.  Christen details the  attacks---dictionary and frequency---that result in hashing not being secure in that case. 

Vatsalan \etal \cite{vatsalan2013taxonomy} provide a taxonomy and overview of the field as a whole. Their taxonomy is both thorough and extensive. It reveals that all the schemes included in their taxonomy, which do not rely in some part on Secure Multi-party Computation (SMC), are susceptible to some form of privacy attack (dictionary, frequency, cryptanalysis). Such a view is not new, Trepetin's review of Privacy Preserving String Comparisons in 2008~\cite{trepetin2008privacy} drew similar conclusions. 

This section is divided into a number of sub-sections, Section~\ref{sec:lit:hashing} discusses some of the schemes based on hashing and HMACs. Section~\ref{sec:lit:bloom} looks at the usage of Bloom filters, whilst Section~\ref{sec:lit:smc} explores Secure Multi-Party Computation approaches. Section~\ref{sec:lit:ons} provides an in-depth analysis of the UK Office of National Statistics approach, which is closely aligned in aims, but unfortunately is  insecure. 

In terms of security the most promising approaches are based on Secure Multi-party Computation (SMC), which is rooted in the more formalised and rigorous field of cryptography. Such schemes offer stricter security guarantees and the potential to perform linking across multiple organisations. However, the current crop of SMC schemes are not efficient enough for deployment at large scale. Additionally, their dependence on having multiple independent parties is at odds with the setting the ABS operates in, in which it is the sole organisation performing the linking. A move to a more distributed, multi-party setting offers significant potential, and should be considered as part of a longer-term strategy. In the interim a more efficient approach will be required to meet the immediate privacy and linking requirements of the ABS.

\subsection[Data Matching: Concepts and Techniques For Record Linkage,Entity Resolution, and Duplicate Detection - Peter Christen]{Data Matching: Concepts and Techniques For Record Linkage,Entity Resolution, and Duplicate Detection - Peter Christen~\cite{christen2012data}}\label{sec:lit:christen}
One of the seminal references in the field is the Data Matching book~\cite{christen2012data} by Peter Christen. The book covers record linkage and the associated methodologies. The Chapter of particular interest for us is \emph{Chapter~8 - Privacy Aspects of Data Matching}. Many of the approaches referenced by Christen in Chapter~8 are explored in more detail below. This section will act as an overview of the chapter as whole. 

\subsubsection{Framing Privacy Preserving Data Matching}\label{sec:christen:framing}

The chapter highlights that damage can occur where identification of only a few, or even just one, record occurs. For example, identifying a politician/celebrity in a medical dataset. Likewise, incorrect matches can cause damage if the output dataset is revealing a particular threat or characteristic, for example, identifying possible terrorists. Christen discusses the wider risks of re-identification that occur, even when identifiers have been removed. De-identification is a topic in its own right, and is beyond the scope of this document, but is of relevance to the data held by the ABS and the ability to re-identify after the deletion of the name and address.  Christen provides an overview of scenarios where privacy and data mining interests have collided and emphasises the importance of addressing it. 

\subsubsection{Trust and Security Models}
In addition to the assumption of honesty in the single organisation setting, there are further assumptions for the multiple organisation setting. Christen provides two categories for the multi-party setting: the three party case, and the two party case. In the three party case a third party is required to perform the linking. In the two party case, two database owners work together, without any additional party, to perform the linking. The three party cases is not ideal due to the need to add an additional party, due to the risks associated with another linking party colluding with, or compromising, the third party. Whilst the two party setup is preferred, it often comes at the cost of higher computation complexity and runtimes. 

Christen provides an overview of two most common security models used in the literature. The first is the  semi-honest model, which is often referred to as the honest-but-curious model. In such a model the participants behave honestly with regard to the protocol, in that they send valid messages and respond accordingly, but they are free to record and infer as much information as they can. As such, they are free to mount frequency attacks and dictionary attacks as we discussed in Section~\ref{sec:hashing} and~\ref{sec:freq}. Participants are also free to augment their knowledge with any publicly available data, for example, the phone book. 

The second security model is the malicious model, which does not place any restrictions on the participants. Furthermore, it allows participants to both deviate from the protocol and send malicious content in order to try and infer further information. It is a much stronger model than the honest-but-curious model, and is used in many secure multi-party computation protocols. Achieving privacy in the malicious model is much harder, due to the power of the adversary, and as such, it often requires the use of sophisticated cryptography. 

\subsubsection{Exact Matching}
Christen explains at length the usage of hashing for exact matching, as well as referencing a number of papers related to schemes developed by French researchers in the 90's. Christen correctly notes the susceptibility of hashing approaches to dictionary attacks. He proceeds to state that due to this weakness ``...simple hash-encoding approaches to privacy-preserving matching can only work for three-party protocols"~\cite{christen2012data}. However, the subsequent paragraph explains that the dictionary attack weakness still exists with three parties, since the third party can mount the dictionary attack itself. The suggested counter to this is to use a keyed hash since this will prevent the dictionary attacks. This is then shown to also not work, since the third party can still mount a frequency attack. As such, there is no demonstration of any protocol that works under either of the defined security models, for either the two or three party setting. This is not explicitly stated, but the consequence is that schemes that utilise such approaches are in fact only sound under a fully-trusted third-party security model. In such a model, the third party is assumed to behave completely honestly and does not try to learn anything beyond what it has been given. 
This is an extremely weak security model which is not appropriate for ABS because it requires complete protection of all the data.

Christen references a number of protocols, mostly secure multi-party protocols, for database record matching and extraction. None of which are relevant to our setting due to their only being able to achieve exact matches. There are also further limitations in that one party potentially learns the contents of a match, which is not consistent with our requirements. 

The summary of the section on hashing states ``...while simple one-way hash encoding allows efficient privacy preserving data matching across organisations, the main limitation of these approaches is that they can only find exact matches"~\cite{christen2012data}. In fact, the main limitation is that it does not provide privacy under either security model defined earlier in the chapter. The fact it only permits exact matching is secondary if an intended privacy preserving record linkage method does not provide privacy. 

\subsubsection{Approximate Matching}
Christen provides short overviews of a number of techniques and challenges in approximate matching, including \cite{atallah2003secure}, \cite{churches2004some}, \cite{schnell2009privacy}, and \cite{kuzu2011constraint}, all of which we cover in more detail below. He also references a number of survey papers, including \cite{durham2012quantifying} and \cite{trepetin2008privacy}, both of which we discuss later in this section. 

The conclusion taken from this part of the chapter, and the subject of the subsequent section, was that schemes did not take into consideration the computation cost of deploying them at scale. This is discussed further in ``A taxonomy of privacy-preserving record linkage techniques"~\cite{vatsalan2013taxonomy}, of which Christen is a co-author. The chapter continues by discussing blocking and indexing techniques that can be used to improve the efficiency of running complex protocols on large datasets. Such approaches will be important to the efficiency of any approach taken by the ABS, but are beyond the scope of this paper.

\subsubsection{Overall}
There is no single recommended approach, with trade-offs present in all. Sometimes those trade-offs are privacy related, sometimes complexity related. The chapter provides suggested areas of further reading, many of which we will refer to in the following sections. The overall conclusion should be that this is still an active area of research, and as such, no definitive solution has been found. 

\subsection{Hashing and HMACs}\label{sec:lit:hashing}
The usage of hashing in privacy preserving record linkage has a long history, and surprisingly, given its weaknesses, is still being used today. Quantin \etal~\cite{quantin2014epidemiological} utilise a salted hash algorithm, based on the approach by Benhamiche and Faivre~\cite{benhamiche1998automatic}, from 1998. The approach combines a cryptographic hash with a salt. A salt is an additional random input combined with the plaintext to prevent simple dictionary attacks. It is used most typically when hashing passwords. Benhamiche and Faivre~\cite{benhamiche1998automatic} use a deterministic salt generation algorithm of their own creation. As a result, the input will always hash to the same output. As a result the output remains susceptible to frequency attacks. Additionally, the usage of a deterministic salt generation algorithm is concerning, as it significantly reduces randomness, and when faced with frequency attacks, could allow recovery of all or parts of the salt codebook. Such an approach is not applicable to our setting because of the susceptibility to frequency attacks and questionable security properties of the underlying primitive. 

\begin{sloppypar}
Weber \etal~\cite{weber2012simple} proposed a similar construction to Quantin \etal~\cite{quantin2014epidemiological}, using a salted MD5 hash. Their contribution was in the construction of a linkage key, which involved an approach similar to the one taken by the ONS~\cite{ONSM9}, except Weber \etal propose just a single linkage key. They postulate that the most errors in names occur after the first two characters. As such, taking the first two characters of the first and last names, and combining them with the date of birth, produces a highly unique linkage key. Whilst they do not provide an in-depth analysis of their assumption, it would appear to be consistent with what is shown by the ONS~\cite{ONSM9}. Furthermore, if the approach does yield highly unique linking keys it will mitigate the chances of a frequency attack on the output tags. They also assume that the date of birth is highly reliable, which could well be true in medical settings, but may not be true on survey/census data. It would seem likely that different formats for dates could easily lead to month and day being transposed. Their approach does not handle fuzzy matching, and there is an inherent reversibility if one has knowledge of the key.
\end{sloppypar}

Churches and Christen~\cite{churches2004some} first evaluate a previously proposed french scheme~\cite{dusserre1994one}, highlighting some weaknesses before proposing a new approach. The paper discusses the weaknesses of Hashes, as well as the weakness that HMACs remains susceptible to frequency attacks. Despite this observation, their proposed scheme still utilises HMACs. The authors justify this on the basis of a trust model that involves trusting a third party to not perform an attack or to collude with another party. The full protocol involves five parties, with various different trust assumptions. There is also a suggestion that having multiple trusted third parties, selecting which one to use at the last minute, will improve security. This assertion is not justified by rigorous security analysis. The approach is not applicable to our setting because the security stems from the trust assumptions, not the protocol or methods themselves. 

Other schemes mix hashing with crytpo via more complicated protocols. For example, Guesdon  \etal~\cite{guesdon2016securizing} apply multiple keyed hashes, as well as using asymmetric cryptography. The cryptography is only used for the transmission of the data, not in the underlying security of it. The scheme relies on a trusted third party that can decrypt the asymmetric ciphers. The scheme also utilises two rounds of keyed hashing, although it isn't clear as to what additional security this really delivers. The scheme remains susceptible to frequency attacks, and combined with dependence on a trusted third party makes it unsuitable for our context. 

Any scheme that relies on directly hashing or HMAC'ing the identifier will not be applicable due to susceptibility to frequency attack. Additionally, despite many papers misunderstanding the nature of hashes, they remain reversible when used over a small input set---the case in our context. HMACs do not solve that problem, they just place a dependence on knowledge of a secret key. As such, they become equivalent to symmetric encryption. If symmetric encryption is deemed inappropriate due to the recoverability, then HMACs are equally inappropriate.

\subsection{Similarity Tables}\label{sec:lit:simtabs}
The approach taken by Pang and Hansen~\cite{pang2006improved} claims to use encryption, although it appears it actually uses a keyed hash function. The misuses and substitution of cryptographic terminology is not uncommon in the field, but presents significant problems when performing a security analysis. The underlying approach is not of particular interest, as it suffers from the same weaknesses as the previously discussed approaches. However, it presents an approach for performing fuzzy matching on the values. The authors propose using a similarity table to calculate string similarity in advance of the linking. 

A similarity table is formed by first constructing a list of distinct values. This could come from the data itself, as in the ONS case~\cite{ONSM9}, or from an independent source, for example a voter registration database or phone book, as suggested by Pang and Hansen~\cite{pang2006improved}. Both parties must have the same list and prior to hashing they calculate the similarity of the plaintexts. If the similarities are below a threshold they are kept and stored in the similarity table, indexed by the hashed identifiers. When performing the linking, non exact matches can be be found by checking for records for the two hashes in the similarity table and retrieving the respective similarity scores. 

The Pang and Hansen protocol~\cite{pang2006improved} is dependent on having two independent parties and a trusted third party to perform the actual linking, and as such is not applicable to our context. However, the concept of pre-calculating a similarity matrix has some merit, and  is used in the ONS~\cite{ONSM9} approach. Whilst the Pang and Hansen~\cite{pang2006improved} approach is not directly applicable, the concept of a similarity table is worth further consideration.

Vatsalan \etal~\cite{vatsalan2011efficient} propose a similar approach although they pre-process the database using phonetic similarity and a blocking process prior to calculating the similarity. Each block is only compared for similarity with the reference values associated with its block, thus reducing the size of the similarity table that needs to be calculated. The authors also use the reverse triangular inequality, instead of the triangular inequality used by Pang and Hansen~\cite{pang2006improved}. Whilst the approach offers some efficiency advantages, it does not fundamentally change the privacy properties, or the reliance on a trusted third party. 

\subsection{Security and accuracy of current literature on Bloom Filters} \label{sec:BloomFilters}  \label{sec:lit:bloom}
Bloom filters~\cite{Bloom1970space} have been widely discussed in the privacy preserving record linkage literature. It is important to note that Bloom filters are a technique for performing set comparison, not preserving privacy. A number of papers have shown that Bloom filters remain susceptible to frequency attacks when used directly with plaintexts. However, that does not preclude their use for comparing sets of privacy-preserving elements. For example, it would be possible to create a Bloom filter from the linking identifiers in Section \ref{hmac_linking_keys}. This would be advantageous if there was a constraint on storage space. However, this would have a significant impact on the efficiency of linking because the efficient indexes that could be created for the HMAC identifiers would no longer be possible. Like other probabilistic linking techniques, the comparison of Bloom filters requires a full cross comparison of the entire database, and for each comparison a full traversal of bits in the respective Bloom filters. This would make full population linking infeasible, and would be computationally expensive in comparison to deterministic linking, even with appropriate blocking. 

In summary, Bloom filters are useful for set comparison, but  are not a privacy preserving method in their own right. 

\subsubsection{Bloom Filter construction}
Constructing an efficient Bloom filter is non-trivial, requiring multiple independent hashes. The simple approach to generating indepedent hashes is to use Universal Hashing in the form of a Carter-Wegman hash. Such hashes are of the form:

$$h_{(a,b)}=((ax+b) \bmod p \bmod L$$
where $a$ and $b$ are random integers $\bmod\  p$,  $a\neq0$ and $L$ is the length of the Bloom filter, which must be a large prime\footnote{Setting $p=L$, both prime, also works without requiring $p$ to be large.  The requirement for $p$ to be large produces a near-uniform distribution for any value of $L$ including composites.}. Independent hashes can be constructed by independently constructing $a_i$, $b_i$,
for $i = 0$ to $k-1$. This was shown by \cite{Ramakrishna:1989} to provide a false positive result close to that of the theoretical optimum. However, it is considered a costly operation to construct and use such hashes, particularly if $k$ is large. 

 Kirsch  {\it et al.} \cite{kirsch2006less} introduced a technique for efficiently constructing hashes from just two independent cryptographic hashes. They provide a proof that their proposal is asymptotically no worse than using $k$ independent hashes. Their construction is of the form 
 
 $$g_i(x)=h_1(x) + ih_2(x) + f(i)~mod~m$$
 
 Where $h_1$ and $h_2$ are independent cryptographic hash functions, for example, SHA1 and MD5.\footnote{SHA1 and MD5 appear in the literature extensively, however, both are considered deprecated as cryptographic hash functions, particularly MD5. It is not immediately clear that using SHA512 and SHA256 would constitute two independent hashes.} If the arbitrary function $f(i)\equiv0$ the schemed is referred to as double hashing, otherwise it is referred to as extended double hashing. The analysis in \cite{kirsch2006less} is with regards to enhanced double hashing. 
 
 The guarantee of performance ``asymptotically no worse'' than the original construction may not mean much in practice.  In this section we investigate  empirically how much worse it is in practice.  We find that many constructions have very high false-positive rates, especially when combined with approximate matching.
 
 An important requirement,  often overlooked, is that $m$ must be prime. This detailed further by Dillinger and Manolios \cite{dillinger2004Bloom}, but is not included in the papers by either Schnell {\it et al.} \cite{schnell2009privacy} or Randall {\it et al.} \cite{randall2014privacy}.
 
 \subsubsection{Bloom filter analysis} \label{subsec:BloomFilterAnalysis}
 
 Bloom filters were originally designed to efficiently determine whether an element was in a set or not. Their use has been expanded, particularly in the privacy preserving record linkage field, to be used as a measure of similarity. This is typically done by calculating the dice-coefficient between two Bloom filters  \cite{schnell2009privacy},\cite{randall2014privacy}. However, there is little theoretical analysis of this usage of Bloom filters. Broder and Mitzenmacher \cite{broder2004network} describe estimating the set intersection between two Bloom filters, however, it is considerably more complicated than a simple dice-coefficient. 
 
We examine below the error rate for Bloom filters, both the original construction and the double hashing version, when combined with an n-gram approach for fuzzy matching of names.  We find very high similarity rates even among data that should not be similar.  These  represent a combination of false matches caused by the n-gram treatment itself,   the imperfect implementation of Bloom filters, and the difficulty of accurately computing similarity measures for Bloom filters.  A typical example is given in Figure~\ref{tab:dh1003largestMain}, which shows similarity scores from using the dice-coefficient on n-grams stored in a Bloom filter using double-hashing.

\begin{table}[htbp]
\begin{center}
\begin{tabular}{|l|l|l|}
\hline
\multicolumn{1}{|c|}{\textbf{Dice-coefficient}} & \multicolumn{1}{c|}{\textbf{Name One}} & \multicolumn{1}{c|}{\textbf{Name Two}} \\ \hline
0.7347 & blocklar & sahinagic \\ \hline
0.7458 & frankenfeld & dhumatkar \\ \hline
0.7429 & kolin & wendly \\ \hline
0.7429 & lehky & turcon \\ \hline
0.7368 & quinney & foica \\ \hline
0.7407 & runt & meij \\ \hline
\end{tabular}
\end{center}
\caption{Double Hashing Bloom Filter to Bi-gram Comparison: high similarity scores for dissimilar names}
\label{tab:dh1003largestMain}
\end{table}

Bloom filters were designed to efficiently determine set membership. Using them for similarity measurement requires re-evaluation of the properties and construction. Our initial analysis indicates the optimisation strategies for similarity measurements are different to those for minimising false positive when testing set inclusion. As such the constructions generally do not work in the way they are intended.

\subsubsection{Use of Bloom filters in linking}
Bloom filters provide an alternative approach to similarity tables for fuzzy matching. Schnell \etal~\cite{schnell2009privacy,schnell2010private} propose using hashes and Bloom filters to enable fuzzy matching. The concept itself is not new~\cite{kirsch2006less}, but its use in privacy preserving record linkage was novel. A Bloom filter is an efficient randomised data structure for storing a set of items, and then determining whether subsequent items are present in the set or not. If a match is found it means an item is probably in the set, if it is not found it means it is definitely not in the set. As such, it permits false positives, but not false negatives. Bloom filters have applications in a number of areas of computing, and as such are well studied. 

The approach of Schnell \etal is to use a standard Bloom filter populated using the bi-grams of the identifiers. When performing matching, two Bloom filters are compared by calculating the Dice Coefficient to give a similarity score that can then be used to determine if it should be treated as a match or not. 

Instead of using hashes Schnell \etal suggest using HMACs to avoid simple dictionary attacks. This again results in a dependence on a key, and the addition of a Bloom filter does not significantly impact on the ability to recover plaintexts if the key is known. It was shown by Bachteler \etal~\cite{bachteler2010empirical} that the approach by Schnell \etal~\cite{schnell2009privacy} was more accurate than the schemes proposed by Pang and Hansen~\cite{pang2006improved} and Scannapieco \etal~\cite{scannapieco2007privacy}. However, in the case of Pang and Hansen~\cite{pang2006improved}, the most significant impact on performance was the selection of a reference table: where it was not a superset of both parties' sets, the accuracy deteriorated significantly. In the ONS~\cite{ONSM9} approach they utilise the superset and report obtaining good results. 

More recently it has been shown by Kuzu \etal~\cite{kuzu2011constraint} that Bloom filters remain susceptible to frequency attacks. Kuzu \etal discuss approaches for making such attacks more difficult, but there is no rigorous proof of security, and as such, counter measures remain ad-hoc and potentially unreliable. 

Bellovin \etal~\cite{bellovin2004privacy} proposed using encrypted Bloom filters for privacy enhanced searching of a database. Their approach involved replacing the hash functions with encryption schemes. The cryptography they used is fairly esoteric, and there appears to be a reliance on hashing as well. They rely on a trusted third party to transform ciphers between different keys. The dependence on a trusted third party and inherent recoverability leads up to conclude that this is not applicable to our context. 

Kerschbaum~\cite{kerschbaum2011public} proposes using public key encrypted Bloom filters for supply chain integrity. The scheme is an interesting one, utilising the underlying homomorphic properties of the selected encryption scheme to protect the contents of the Bloom filter. The scheme utilises zero-knowledge proofs to enable non-interactive operations to be performed in a verifiable manner. The scheme relies on advanced cryptography, notably the Goldwasser-Micali (GM) encryption scheme. It is not immediately clear that the approach could be modified to enable fuzzy matching, due to the Bloom filter comparison being performed in the encrypted domain, which restricts the ability to calculate a similarity score like the Dice coefficient.

\subsubsection{Uniformity of output}
\begin{figure}[htbp]
\centering
\includegraphics[width=0.8\textwidth]{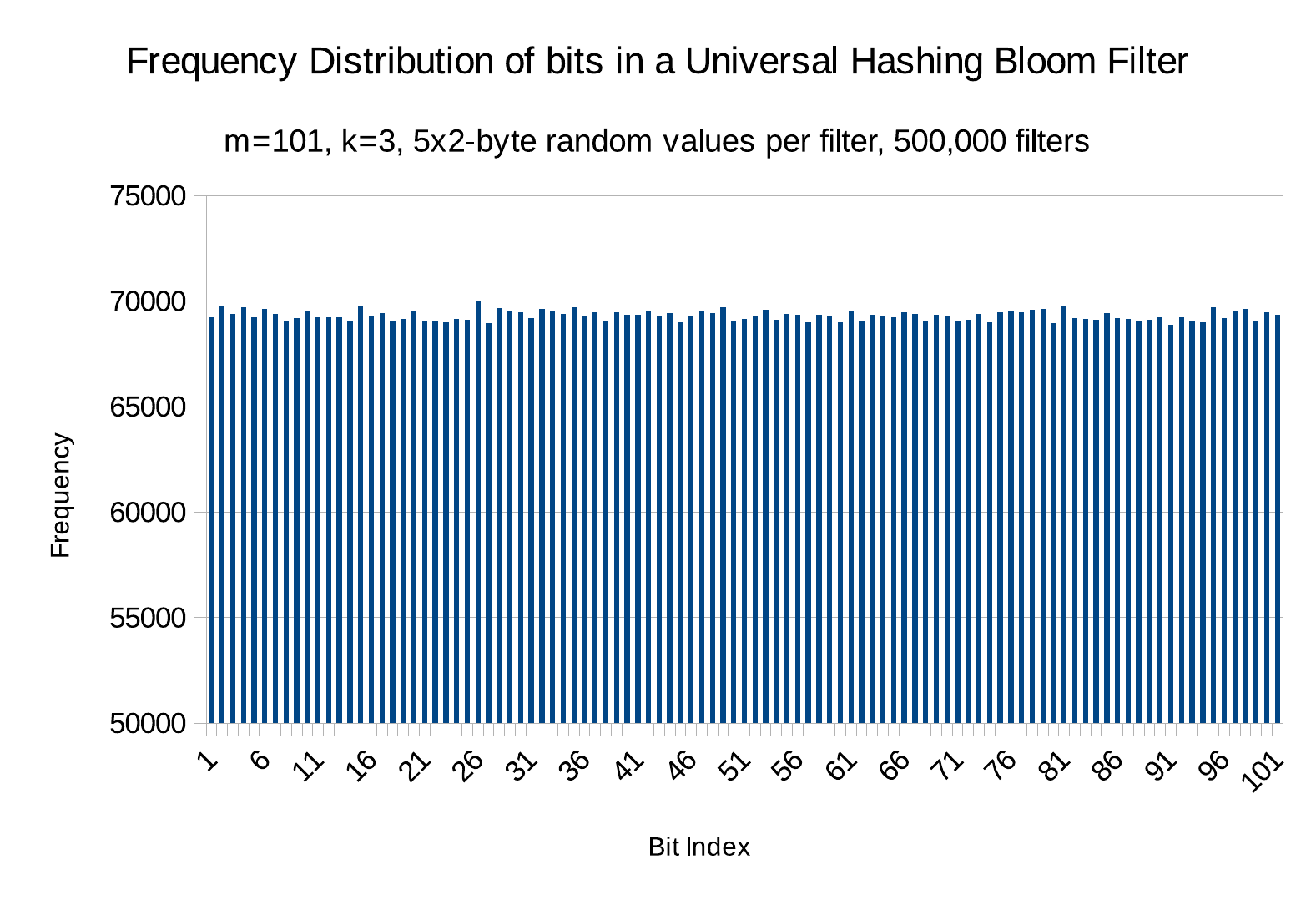}
\caption{Frequency Distribution of Universal Hash based Bloom Filter}
\label{fig:uh101}
\end{figure}

This section examines the use of the dice-coefficient as a method of estimating
the similarity of two sets of elements in a Bloom filter.
 One of the implicit aspects of using a dice-coefficient is that the bits being compared are of equal weight. As such, there is an implicit assumption that the Bloom filter is uniformly distributed. Were that not to be the case the bits with a higher frequency could distort the similarity scoring. If we look at the ideal implementation of a Bloom filter, using Universal Hashing, we can see it produces a largely uniform output---see Figure~\ref{fig:uh101}. 
 
Figure \ref{fig:uh101} was generated by constructing 500,000 Bloom filters based on 3 independently generated Universal Hashing algorithms. Into each Bloom filter 5 randomly generated 2-byte values were submitted, to simulate bi-gram insertion. All 500,000 Bloom filters were then compared to calculate the frequency of each bit being set to 1. This uniform output is exactly what we would expect to see. 

\begin{figure}[htbp]
\centering
\includegraphics[width=0.8\textwidth]{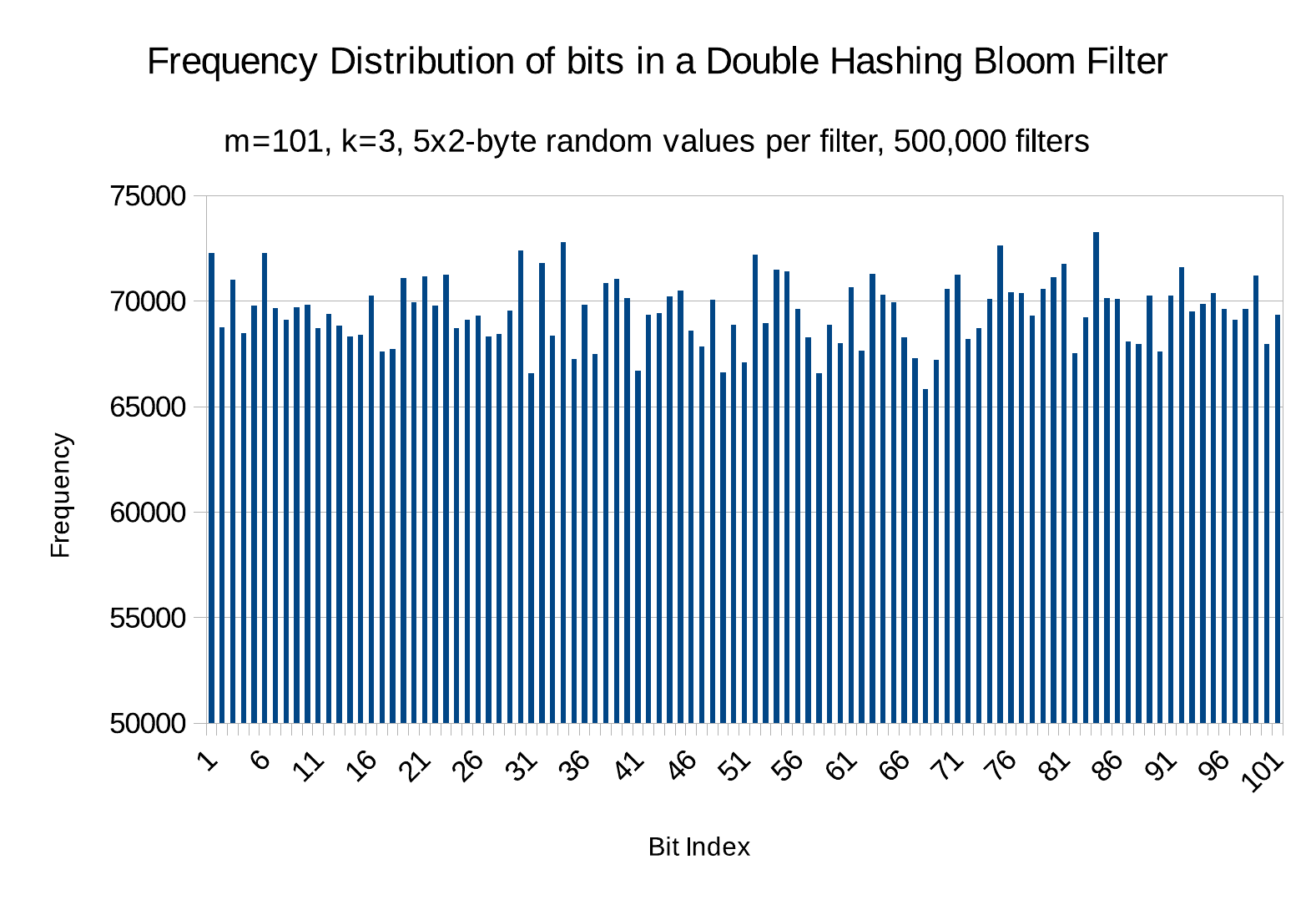}
\caption{Frequency Distribution of Double Hash based Bloom Filter}
\label{fig:dh101}
\end{figure}

In contrast, Figure \ref{fig:dh101} shows the frequency distribution of double hashing based Bloom filters. These filters were based on SHA1 and MD5 \cite{randall2014privacy}\cite{schnell2009privacy}, using a size of 101 and $k=3$. A size of 101 was selected to be the closet prime to the size used in \cite{randall2014privacy}. The same approach of generating 500,000 Bloom filters, each with 5 random 2 byte values submitted to them. It is clear from Figure \ref{fig:dh101} that there is less uniformity in this construction. This is further evidenced by the data in Figure \ref{fig:uh101} having a standard deviation of 233, whilst the data in Figure \ref{fig:dh101} has a standard deviation of 1570. However, it is not immediately clear what impact this will have on the dice-coefficient. We  evaluate that in the Section \ref{Bloom:similarity}. 

\subsubsection{Use of random values}\label{sec:randvalues_bigramskew}
The reason we evaluated uniformity using random values instead of data from our synthetic dataset is due to the non-uniform distribution of the bi-grams from last names. A skewed input set will result in a skewed output set, hence we wanted to eliminate that from our evaluation. In Section \ref{Bloom:similarity} we evaluate against our synthetic dataset, since the evaluation of similarity in names is the specific application we are evaluating, and the skew in the input set will be present in any deployment, so should be considered something that the approach must be able to handle. 

\begin{figure}[htbp]
\centering
\includegraphics[width=0.8\textwidth]{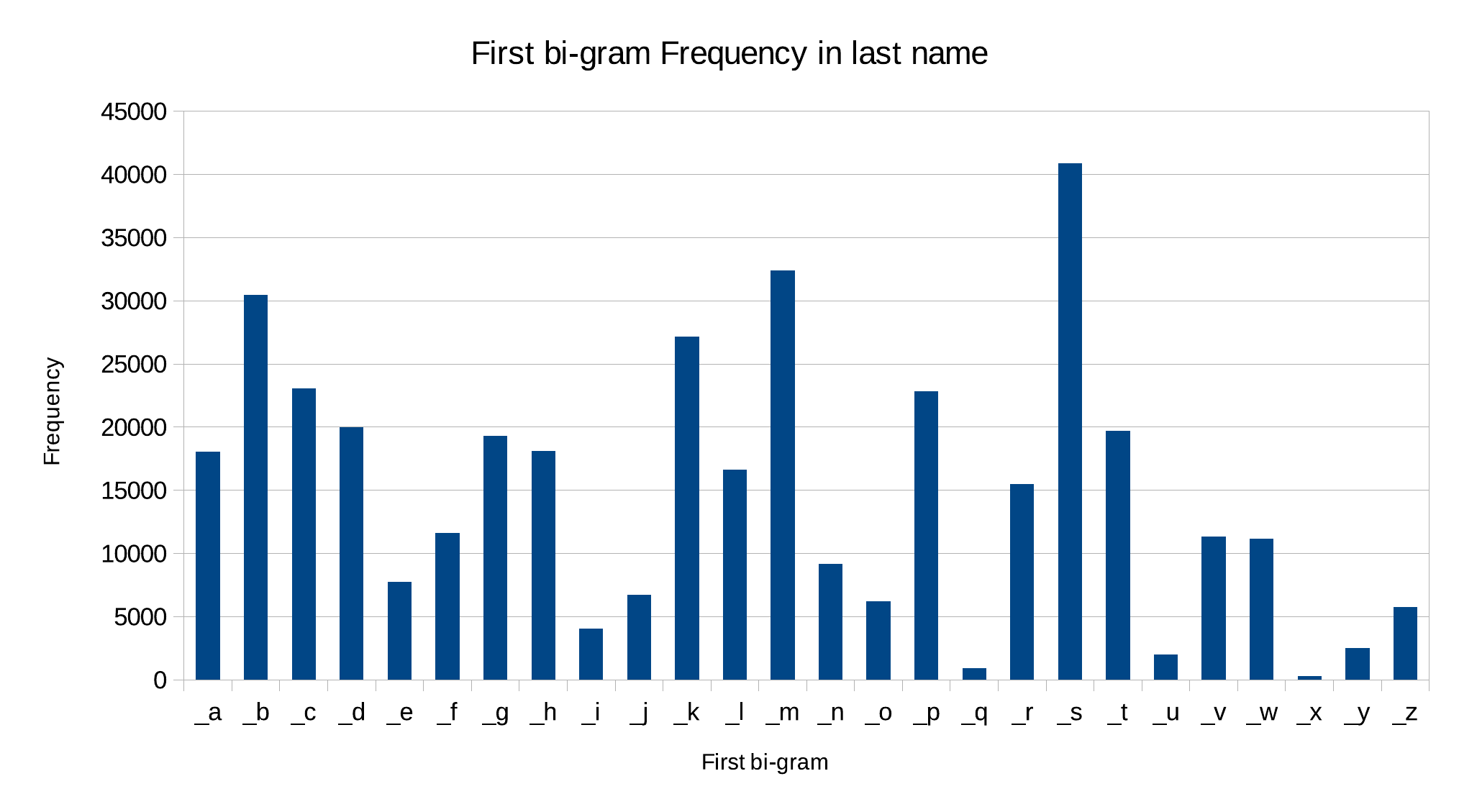}
\caption{Frequency Distribution of First Bi-Gram in Last Names}
\label{fig:firstbilastname}
\end{figure}

As an example of the skew of the input set, Figure \ref{fig:firstbilastname} shows the frequency distribution of the first bi-gram in the last name dataset we have, which consists of 384,370 real last names found in Australia. It is clear to see the prevalence of last names beginning with the letter ``s", and the extreme rarity of ``q" and ``x".\footnote{Last names were were converted to lowercase to not adversely weight the first letter}

\begin{figure}[htbp]
\centering
\includegraphics[width=0.8\textwidth]{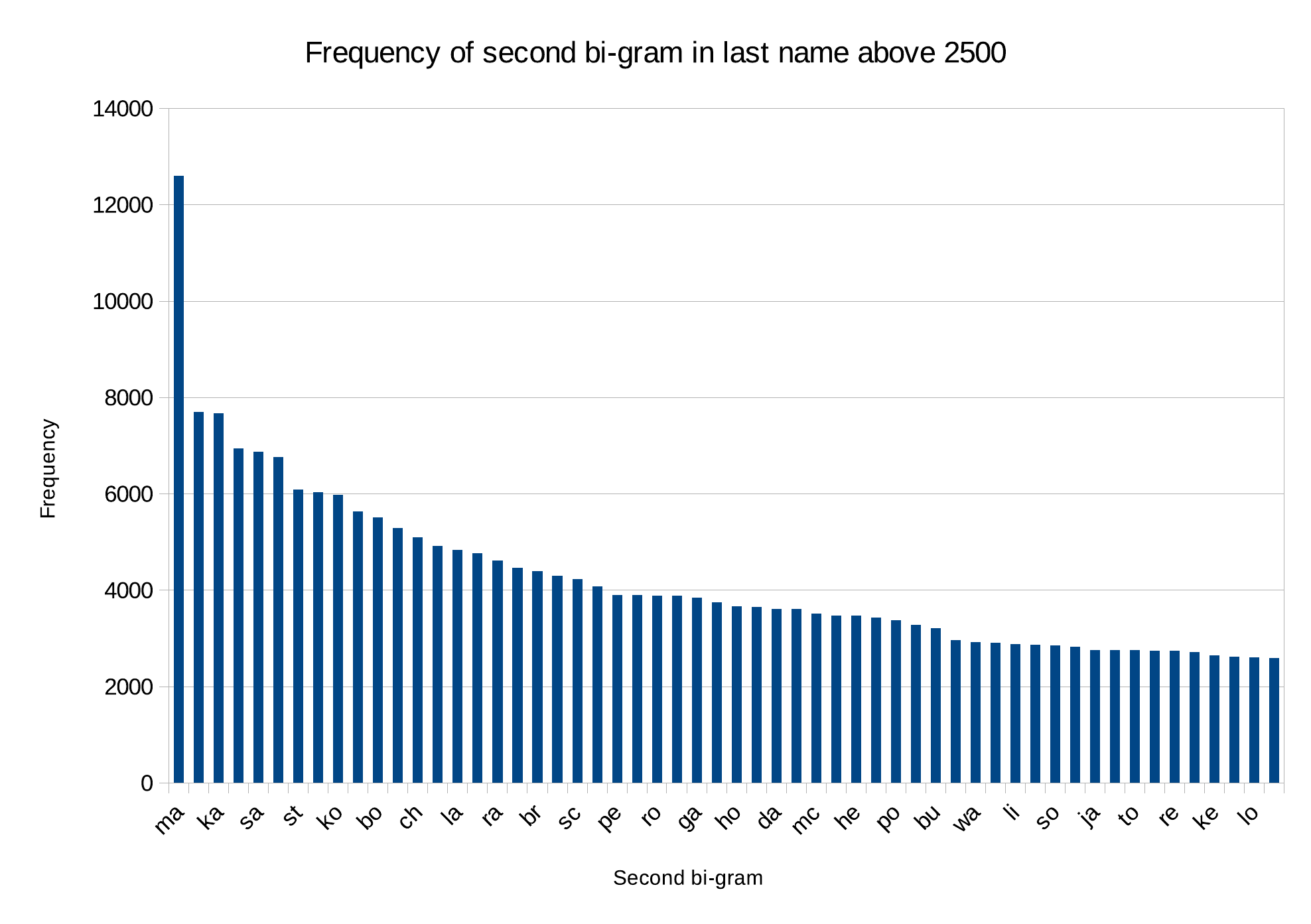}
\caption{Frequency Distribution of Second Bi-Gram in Last Names (over 2\,500)}
\label{fig:secondbilastname}
\end{figure}

A different trend is continues with the second bi-gram, although there is a much wider range of possible values. Figure \ref{fig:secondbilastname} shows the distribution for the second bi-grams, for bi-grams that had a frequency above 2,500. This threshold was selected to allow the graph to remain readable, and to show a representation of the shape of the frequency distribution. The distribution is ordered by size.

With such skewing in the input set we would expect to see some degree of skew in the output set as well. 

\subsubsection{Similarity score evaluation}\label{Bloom:similarity}
Whilst the application of Bloom filters in the general case is of interest, we are primarily of interest in how it performs in evaluating the similarity of names, and thus, we will focus on actual name data in this section. Our aim was to compare plaintext comparisons of bi-grams with comparisons bi-grams encoded into Bloom filters. In order to do this we took our dataset of 384,370 distinct real last names and constructed bi-grams for each one. For each name we constructed a Bloom filter based on double hashing with a length 100 and 101, both with $k$ of 3, using SHA1 and MD5 to replicate the setup in \cite{randall2014privacy}. We also constructed a Bloom filter based on Universal Hashing, with a size of 101 and $k$ of 3. We then performed a cross comparison of each entry, comparing the Bloom filters with each other, and the plaintext bi-grams. In the case of the plaintext bi-grams the comparison was performed using sets, in that the plaintext bi-grams were added to a set, and the comparison was performed between those sets. This was to ensure a fair comparison, since both comparison would be performed on distinct bi-grams without respect to ordering. In both cases we calculated the dice-coefficient between the Bloom filters, and bi-gram sets respectively. We then compared the similarity score produced by the dice-coefficient to evaluate how similar they were. If the Bloom filter provides an accurate measure of similarity it should produce similar similarity scores to the plaintext set comparison. 

\subsubsection{Loss of ordering}
One of the side-effects of Bloom filters is that they do not maintain order of n-grams, they construct a set of distinct n-grams without regard to the order. As mentioned above we replicated this in our plaintext n-gram comparison to provide a fair comparison. However, the loss of ordering does create a higher risk of false positives.  

\begin{table}[htbp]
\begin{center}
\begin{tabular}{|l|l|l|l|}
\hline
\multicolumn{1}{|c|}{\textbf{Names}} & \textbf{} & \textbf{} & \textbf{Bi-grams} \\ \hline
petitt & pettit & pettitt  & \_p, pe, et, ti, it, tt, t\_ \\ \hline
mamara & marama & maramara &  \_m, ma, am, ar, ra, a\_ \\ \hline
lewellyn & llewellyn & llewelyn & \_l, le, ew, we, el, ll, ly, yn, n\_ \\ \hline
takata & takataka & tataka & \_t, ta, ak, ka, at, a\_ \\ \hline
linemann & linneman & linnemann & \_l, li, in, ne, em, ma, an, nn, n\_  \\ \hline
mulally & mullally & mullaly &  \_m, mu, ul, la, al, ll, ly, y\_ \\ \hline
bebee & beebe & beebee & \_b, be, eb,  ee, e\_ \\ \hline
kirisits & kiritsis & kitsiris &  \_k, ki, ir, ri, is, si, it, ts, s\_ \\ \hline
minisi & minisini & misini & \_m, mi, in, ni, is, si, i\_ \\ \hline
kaparas & karapapas & karapas & \_k, ka, ap, pa, ar, ra, as, s\_ \\ \hline
hanemann & hanneman & hannemann & \_h, ha, ne, em, ma, an, nn, n\_ \\ \hline
amara & amarama & arama & \_a, am, ma, ar, ra, a\_ \\ \hline
pulella & pullela & pullella & \_p, pu, ul, le, el, ll, la, a\_ \\ \hline
debeen & deebeen & deeben & \_d, de, eb, be, ee, en, n\_ \\ \hline
peirrera & pereirra & perreira & \_p, pe, ei, ir, rr, re, er, ra, a\_ \\ \hline
\end{tabular}
\end{center}
\caption{Last names producing identical bi-grams}
\label{tab:identicalbigrams}
\end{table}

Any of the entries in the same row in Table \ref{tab:identicalbigrams} will appear to be exact matches even when they are not. This is largely unavoidable when using Bloom filters,\footnote{Whilst location could be included in the Bloom filter it would significantly impact on calculating similarity. If the Bloom filter was being used as originally designed, for exact matching, it would be prudent to include bi-gram location or to simply submit the name as a whole instead of n-grams. However, for similarity comparison n-grams without location are essential} however, when comparing plaintext n-grams it would be possible to retain ordering and use it in a more sophisticated similarity scoring technique, for example, looking at edit distance. However, such a comparison will significantly increase the computational costs of performing the cross-comparison and it is difficult to see how such an approach could be used in a privacy preserving manner, since recovery of plaintext n-grams is going to leak a significant amount of information.

\subsubsection{Double hashing: Size of 100, \texorpdfstring{$k=3$}{k=3}}
The results when comparing bi-grams of size 100, with $k=3$ are shown in Table \ref{tab:dh1003output}. Table \ref{tab:dh1003output} shows the total number of comparisons made, which is the full cross comparison of our dataset of last names. The table goes on to show the number of those comparisons where the similarity score for the bloom filter was equal to the similarity score calculated on the bi-grams directly. For those comparisons in which the bloom filter similarity score is greater than the direct bi-gram score we also calculate the mean and standard deviation of the difference.

Table \ref{tab:dh1003output} shows that in the vast majority of cases the Bloom filter comparison over estimates the similarity, approximately 97\% of the time. The mean difference, when it does over estimate the similarity, is approximately 0.2, the standard deviation of this difference is 0.0858\footnote{Due to the extremely large size of the dataset the standard deviation is calculated as a running standard deviation using the Welford method}. 

\begin{table}[htbp]
\begin{center}
\begin{tabular}{|l|r|r|}
\hline
 & \multicolumn{1}{c|}{\textbf{Totals}} & \multicolumn{1}{c|}{\textbf{Percentage (\%)}} \\ \hline
\textbf{Total Num. Comparisons} & 147383049024 & 100.00\% \\ \hline
\textbf{Equal Comparisons} & 4478110664 & 0.030 \\ \hline
\textbf{Bloom Filter Greater than n-gram Comp.} & 142904938360 & 0.970 \\ \hline
\textbf{Mean Difference when Bloom Filter Greater} & 0.202 & \multicolumn{1}{l|}{} \\ \hline
\textbf{Standard Deviation of difference} & 0.0858 & \multicolumn{1}{l|}{} \\ \hline
\end{tabular}
\end{center}
\caption{Double Hashing Bloom Filter to Bi-gram Comparison}
\label{tab:dh1003output}
\end{table}

In addition to the overall picture of over estimation, there are also examples of extreme over estimation. Table \ref{tab:dh1003largest} (already shown in Section~\ref{subsec:BloomFilterAnalysis}) shows some of the largest over estimates that were observed across the entire dataset. In this cases the two names do not share any bi-grams, yet they all receive a dice-coefficient on Bloom filter comparison of in excess of 0.73. 

\begin{table}[htbp]
\begin{center}
\begin{tabular}{|l|l|l|}
\hline
\multicolumn{1}{|c|}{\textbf{Dice-coefficient}} & \multicolumn{1}{c|}{\textbf{Name One}} & \multicolumn{1}{c|}{\textbf{Name Two}} \\ \hline
0.7347 & blocklar & sahinagic \\ \hline
0.7458 & frankenfeld & dhumatkar \\ \hline
0.7429 & kolin & wendly \\ \hline
0.7429 & lehky & turcon \\ \hline
0.7368 & quinney & foica \\ \hline
0.7407 & runt & meij \\ \hline
\end{tabular}
\end{center}
\caption{Double Hashing Bloom Filter to Bi-gram Comparison Largest Over Estimation}
\label{tab:dh1003largest}
\end{table}

In addition to showing the over estimation in the Bloom filter comparison, this also provides a good practical example of the infeasibility of large cross comparison. In order to construct the comparisons shown we had to perform 147,383,049,024 comparisons. This less than the square of our dataset size due to the removal of duplicates. In order to process that number of comparisons, on an i7 Quad Core CPU, with all Bloom filters and n-grams pre-computed and held in memory, it took over 6 hours. This is a relatively small cross comparison, approximately 383,904 rows. although it does constitute two comparisons for each row. However, it is easy to see that even if that time was halved, the computational time for performing cross comparisons rapidly grows to the point of being infeasible. 

\subsubsection{Double hashing: Size of 101, \texorpdfstring{$k=3$}{k=3}}
In order to establish that the primary cause of the problem was not just that the size was not prime, we repeated the experiment with the size set to 101. The results are shown in Table \ref{tab:dh1013}. The tables are almost identical, which indicates the underlying problem is not the primality of the size of the Bloom filter. 

\begin{table}[htbp]
\begin{center}
\begin{tabular}{|l|r|r|}
\hline
 & \multicolumn{1}{c|}{\textbf{Totals}} & \multicolumn{1}{c|}{\textbf{Percentage (\%)}} \\ \hline
\textbf{Total Num. Comparisons} & 147383049024 & 100.00\% \\ \hline
\textbf{Equal Comparisons} & 4701012536 & 0.032 \\ \hline
\textbf{Bloom Filter Greater than n-gram Comp.} & 142682036488 & 0.968 \\ \hline
\textbf{Mean Difference when Bloom Filter Greater} & 0.202 & \multicolumn{1}{l|}{} \\ \hline
\textbf{Standard Deviation of difference} & 0.0861 & \multicolumn{1}{l|}{} \\ \hline
\end{tabular}
\end{center}
\caption{Double Hasing Bloom Filter to Bi-gram Comparison with Prime Size (101)}
\label{tab:dh1013}
\end{table}

Again, we see similar results at the extremes with Bloom filters having high dice-coefficients between names that do not share any bi-grams, as shown in Table \ref{tab:dh1013largest}. 

\begin{table}[htbp]
\begin{center}
\begin{tabular}{|r|l|l|}
\hline
\multicolumn{1}{|c|}{\textbf{Dice-coefficient}} & \multicolumn{1}{c|}{\textbf{Name One}} & \multicolumn{1}{c|}{\textbf{Name Two}} \\ \hline
0.7385 & baiden-amissah & tuzzolino \\ \hline
0.7429 & castrechini & prawirohardjo \\ \hline
0.7500 & eykhof & sliz \\ \hline
0.7429 & keays & goeby \\ \hline
0.7586 & lera & pagni \\ \hline
0.7391 & tempalar & sedcole \\ \hline
\end{tabular}
\end{center}
\caption{Double Hashing Bloom Filter to Bi-gram Comparison with Prime Size (101) Largest Over Estimation}
\label{tab:dh1013largest}
\end{table}

\subsubsection{Universal hashing: Size of 101, \texorpdfstring{$k=3$}{k=3}}
If the over estimate observed above is due to the underlying hashing technique we would expect the Universal Hashing to not produce a similar over estimation. However, as we see in Table \ref{tab:cw1013} the same problem exists.

\begin{table}[htbp]
\begin{center}
\begin{tabular}{|l|r|r|}
\hline
 & \multicolumn{1}{c|}{\textbf{Totals}} & \multicolumn{1}{c|}{\textbf{Percentage (\%)}} \\ \hline
\textbf{Total Num. Comparisons} & 147383049024 & 100.00\% \\ \hline
\textbf{Equal Comparisons} & 3983997357 & 0.027 \\ \hline
\textbf{Bloom Filter Greater than n-gram Comp.} & 143399051667 & 0.973 \\ \hline
\textbf{Mean Different when Bloom Filter Greater} & 0.214 & \multicolumn{1}{l|}{} \\ \hline
\textbf{Standard Deviation of difference} & 0.0876 & \multicolumn{1}{l|}{} \\ \hline
\end{tabular}
\end{center}
\caption{Universal Hashing Bloom Filter to Bi-gram Comparison with Prime Size (101)}
\label{tab:cw1013}
\end{table}

Furthermore, as shown in Table \ref{tab:cw1013largest}, we see similar examples of high similarity scores for names which do not share any bi-grams.

\begin{table}[htbp]
\begin{center}
\begin{tabular}{|r|l|l|}
\hline
\multicolumn{1}{|c|}{\textbf{Dice-coefficient}} & \multicolumn{1}{c|}{\textbf{Name One}} & \multicolumn{1}{c|}{\textbf{Name Two}} \\ \hline
0.765 & baban & murati \\ \hline
0.769 & diz & pats \\ \hline
0.744 & himarios & nikezic \\ \hline
0.757 & lebris & ravino \\ \hline
0.750 & singsuk & lebudel \\ \hline
0.750 & waku & arrama \\ \hline
\end{tabular}
\end{center}
\caption{Universal Hashing Bloom Filter to Bi-gram Comparison with Prime Size (101) Largest Over Estimation}
\label{tab:cw1013largest}
\end{table}

\subsubsection{Importance of filter size to \texorpdfstring{$k$}{k}}
The results above indicate that the root cause of the over estimate is not the underlying hash construction. In order to further evaluate how changing the parameters impacted on the over estimate we constructed a random sample of 10,000 last names. Using this smaller sample allowed us to run more cross comparisons and therefore evaluate a larger range of values. We repeated the tests from above to confirm that the sample was an accurate reflection of the larger dataset. 

\begin{figure}[htbp]
\centering
\includegraphics[width=0.8\textwidth]{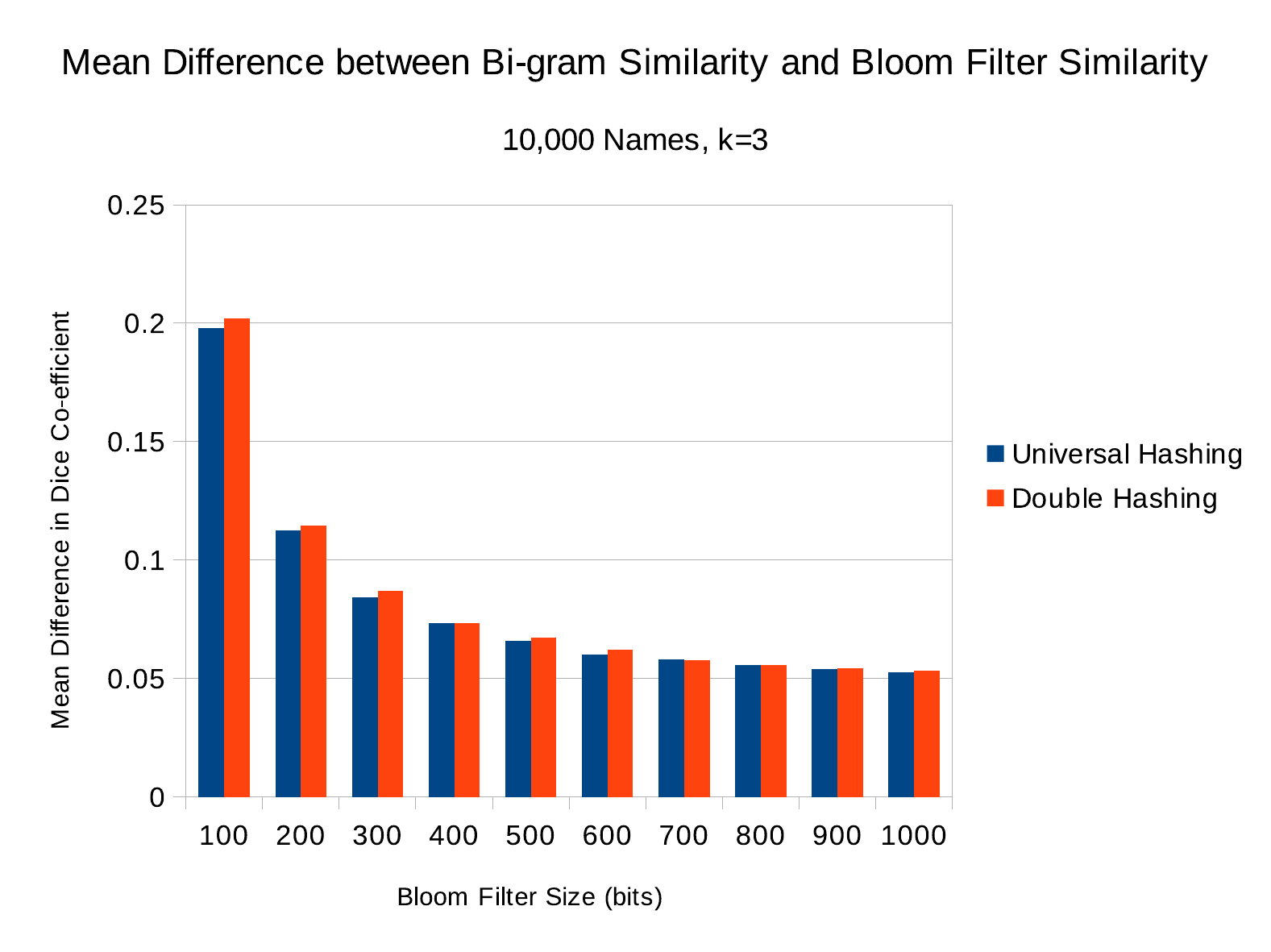}
\caption{Comparison of Similarity Differences as Filter Size Grows}
\label{fig:meandiffk3}
\end{figure}

The first parameter we examined was the size of the Bloom filter. We fixed $k=3$ and increased the size of the Bloom filter in increments of 100 between 100 and 1000. As we can see from the graph in Figure \ref{fig:meandiffk3} as the size of the filter increases the over estimate reduces. Both the Double Hashing approach and the Universal Hashing approach show the same changes. However, it should be noted that whilst the size of the over estimation reduces as the Bloom filter size increases, the rate of reduction also reduces, indicating that there may be some base level of over estimation that cannot eliminated. 

\begin{figure}[htbp]
\centering
\includegraphics[width=0.8\textwidth]{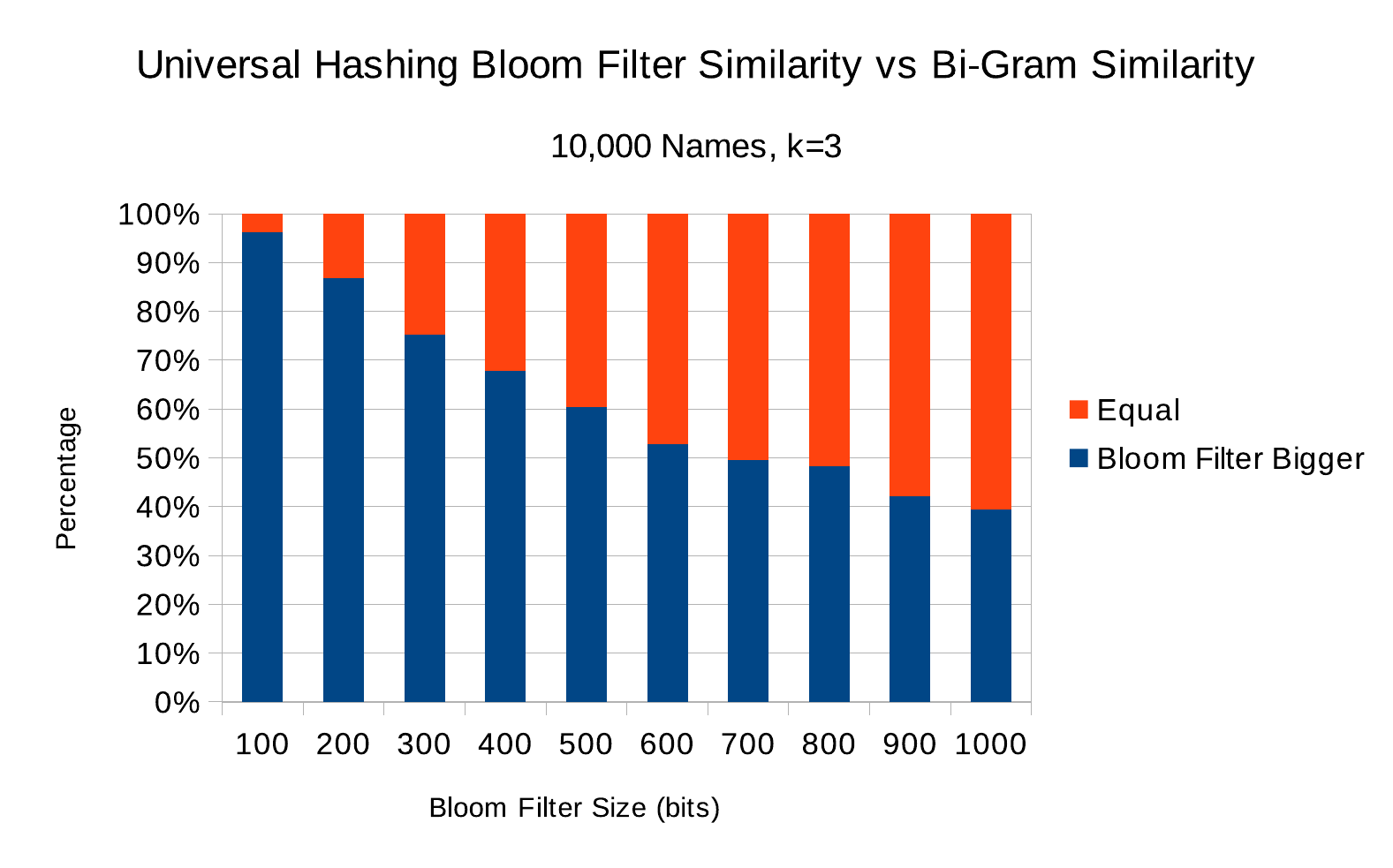}
\caption{Comparison of Percentages of Larger or Equal Similarity Scores}
\label{fig:unihash3diff}
\end{figure}

\begin{figure}[htbp]
\centering
\includegraphics[width=0.8\textwidth]{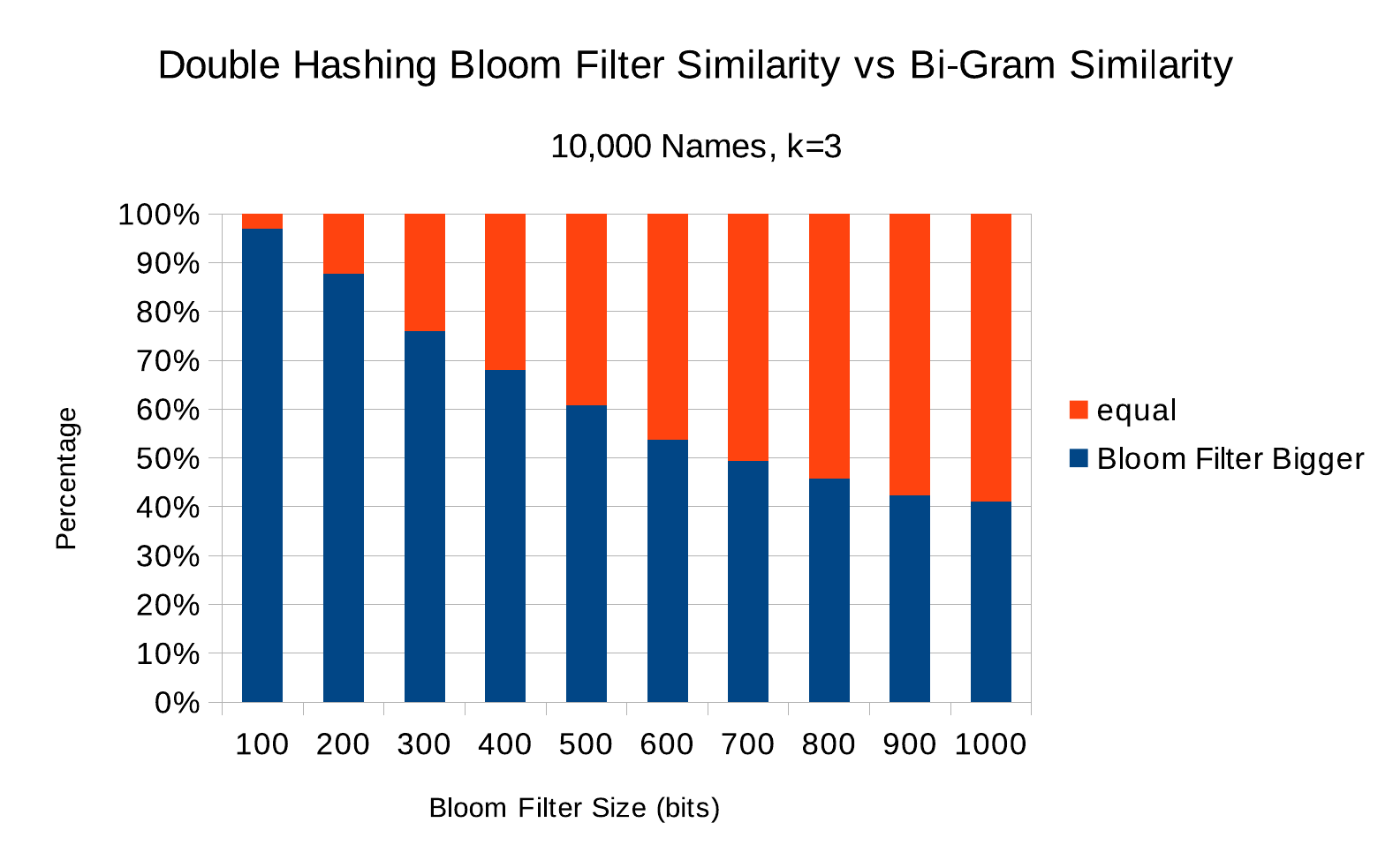}
\caption{Comparison of Similarity Differences as Filter Size Grows}
\label{fig:dblhash3diff}
\end{figure}

In addition to the size of the over estimation reducing, the percentage of comparisons that were over estimated also reduces. Figure \ref{fig:unihash3diff} and \ref{fig:dblhash3diff} shows the graphs comparing the percentage that were equal vs the percentage that were larger, for Universal Hashing and Double Hashing respectively. It should be noted that the n-gram similarity score was never larger.

The reduction in the percentage of comparisons that was larger, in addition to the reduction in the size of the difference is doubly beneficial. However, the graphs indicate that it requires a filter of size 700 bits before the majority are not over estimated. This is a significantly larger Bloom filter, which will impact on performance and storage cost. We have used the value of 10 as the number of bi-grams we expect to put into a Bloom filter, which is broadly inline with our analysis of our last name dataset, which has an average of 9 bi-grams. An estimate of 10 is sufficient for our testing, however, at deployment it may need to be larger, which would have a further negative impact on the efficiency of Bloom filter. 

It would be tempting to assume that it was purely the size of the Bloom filter that is important, but as we shall see, it is the ratio between the size and $k$ the number of hashes that is important. To confirm this we fixed the size at 1000 bits and then increased $k$ in increments of 3 from 3 to 30. 

\begin{figure}[htbp]
\centering
\includegraphics[width=0.8\textwidth]{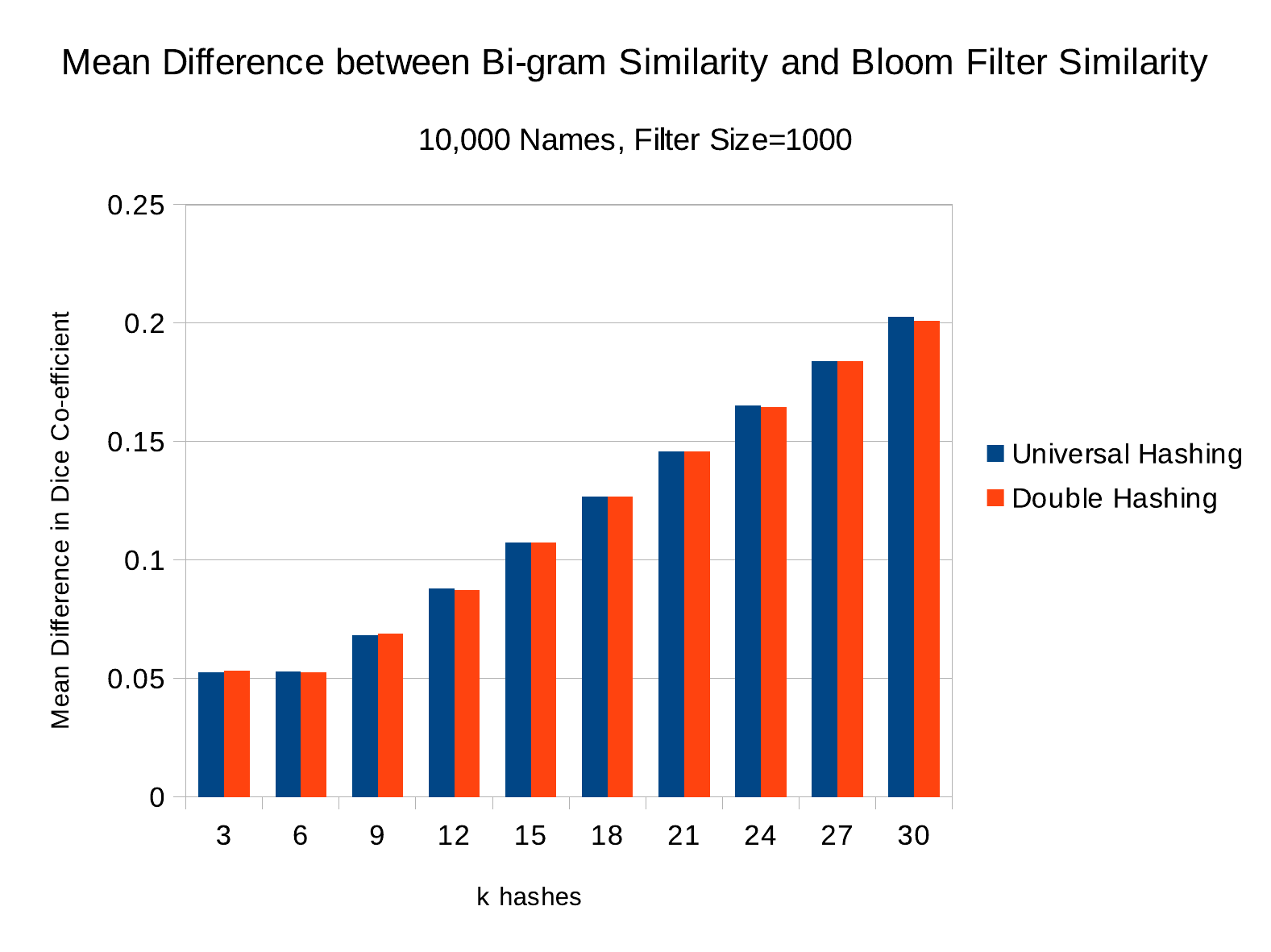}
\caption{Comparison of Similarity Differences as Number of Hashes Grows}
\label{fig:meandiffl1000}
\end{figure}

Figure \ref{fig:meandiffl1000} shows the graph of the results and as expected, it is the inverse of what we saw in Figure \ref{fig:meandiffk3}. As $k$ increases and the ratio reduces the over estimation increases.

This demonstrates that a size of 100 and $k=3$, as suggested by Randall {\it et al.} \cite{randall2014privacy}, produces a high level of over-estimation. Furthermore, the recommendation to maintain the ratio between the size and $k$ as was used in \cite{schnell2009privacy} has the effect of persisting the over estimation. Figure \ref{fig:meandiffl1000} shows that the parameters used by Schnell {\it et al.}\cite{schnell2009privacy} of a size of 1000 and $k=30$ suffers the same over estimation. 

\begin{figure}[htbp]
\centering
\includegraphics[width=0.8\textwidth]{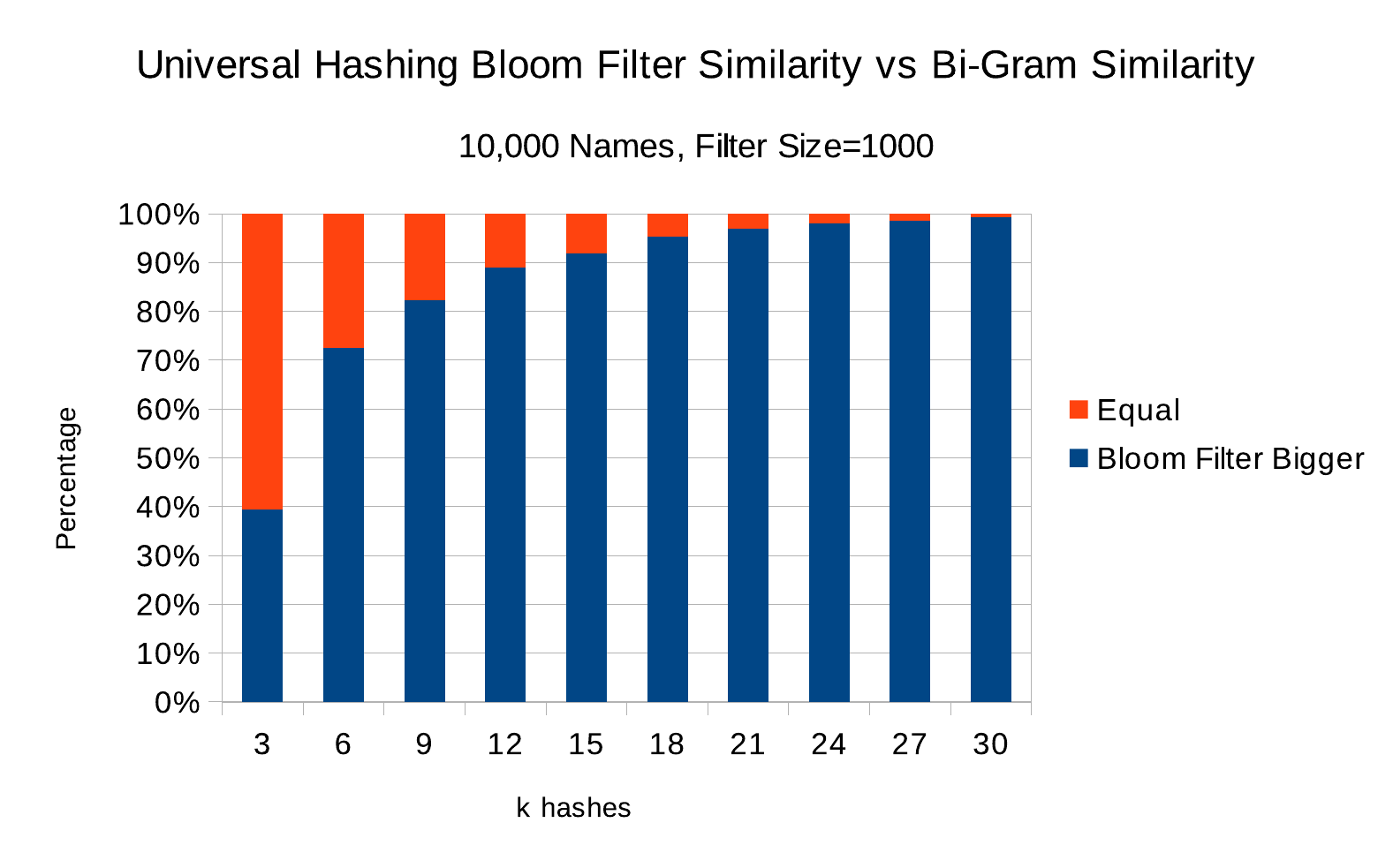}
\caption{Comparison of Percentages of Larger or Equal Similarity Scores}
\label{fig:unihash1000diff}
\end{figure}

\begin{figure}[htbp]
\centering
\includegraphics[width=0.8\textwidth]{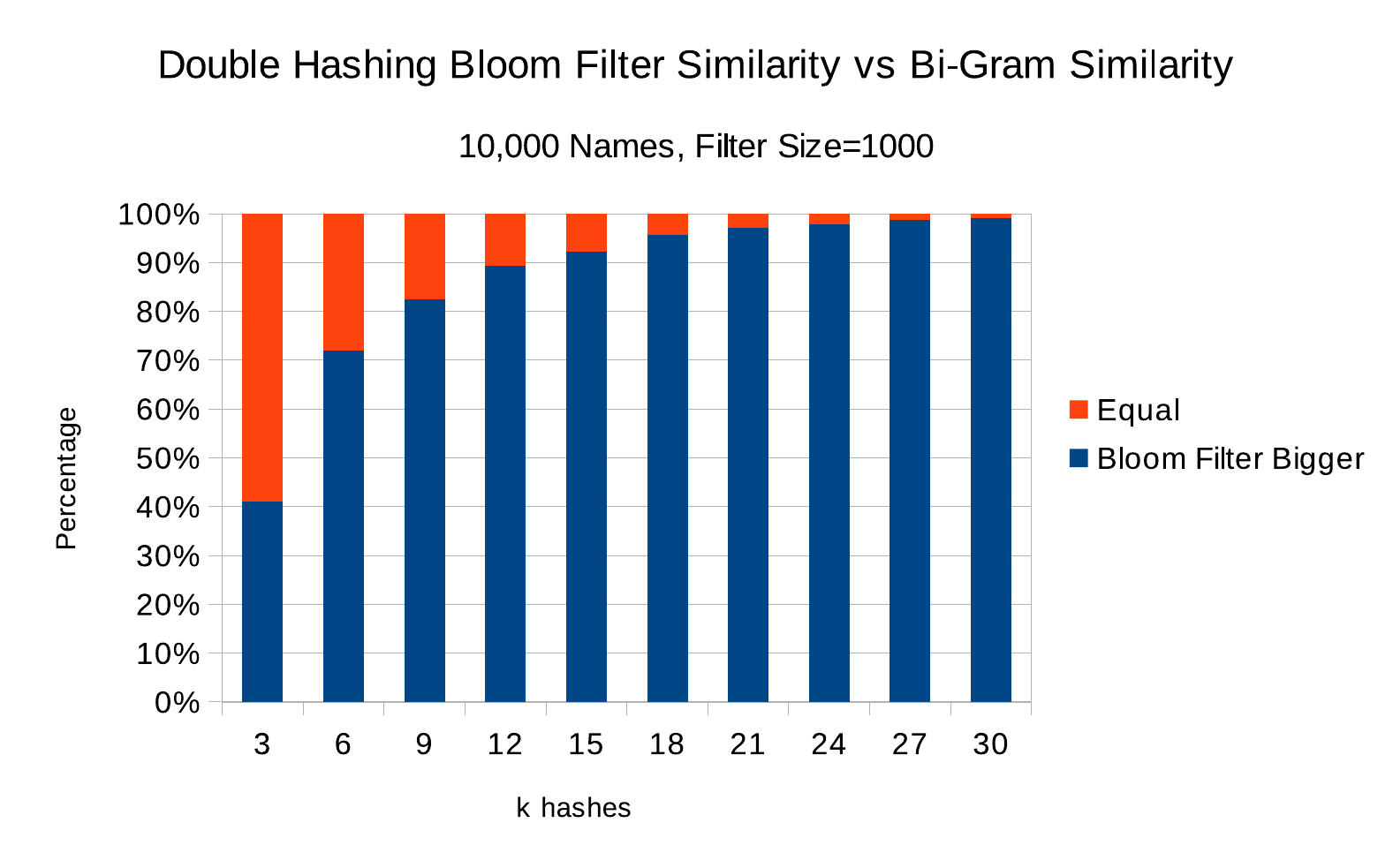}
\caption{Comparison of Similarity Differences as Filter Size Grows}
\label{fig:dblhash1000diff}
\end{figure}

We see the same story with regards to the percentage of comparisons that were over estimated. Figure \ref{fig:unihash1000diff} and \ref{fig:dblhash1000diff} shows the graphs comparing the percentage that were equal vs the percentage that were larger, for Universal Hashing and Double Hashing respectively. 

We can conclude that to minimise both the number and level of over estimate we need a large Bloom filter with a small $k$. This is at odds with the literature which indicates that the ratio of size 100 to $k=3$ should be maintained to achieve results. In our analysis this offers particularly poor performance in terms of the number and scale of over estimation. As the size of the Bloom filter increases it becomes less efficient and more costly to store. As a result, given the limited domain of inputs we have and the relatively small size of the our strings, it is highly likely that straight comparison of bi-grams will give a better performance in terms of speed, storage and accuracy. Additionally, the optimal size of $k$, in terms of false positives is calculated based on the ratio between the number of inputs and the size of the Bloom filter. As such, minimising $k$ with a large Bloom filter could increase the false positive rate. This may not be as important in this instance, since we are only ever performing a similarity score, not a set comparison. However, it highlights the problem of using a Bloom filter which is designed for evaluating set inclusion for measuring similarity. What is optimal for one may not be optimal for the other. 

\DE{what about all with n-grams or n-grams with  some other comparator method}

\subsection{Other Approaches}\label{sec:lit:other}
 Lyons \etal~\cite{lyons2009sail} present their approach to matching records across multiple datasets in the context of the UK National Health Service. Their scheme, SAIL (Secure Anonymised Information Linkage) system has run in the Welsh NHS system. It has been used for exact matching between Anonymous Linking Fields (ALFs) which are derived from the patient identifier via the Blowfish cryptographic algorithm. The approach is not directly related to our context, in that it does not allow fuzzy matching, provides inherent recoverability, and would not protect against frequency attacks. However, it is of note because it is at least using appropriate cryptography for the job it is trying to do, as opposed to relying on hashing, and has been deployed at scale.

Scannapieco \etal~\cite{scannapieco2007privacy} present an approach based on a SparseMap. The concept is to map values into a vector space based on their similarity to a set of randomly selected words. The dimensions of the vector space is reduced, one assumes resulting in a loss of exact matching. The scheme relies on a trusted third party to compare the vector spaces from two parties. The security analysis around what the third party can learn during this process is lacking. It is possible that some information leaks during this process, to the extent that Vatsalan \etal~\cite{vatsalan2013taxonomy} consider it to be susceptible to frequency and collusion attacks. 

Karakasidis \etal~\cite{karakasidis2012fake} propose a scheme that involves injecting fake values into hash value sets to hide the frequency distribution. They do not even propose using a keyed hashing function, instead proposing the use of MD5. The use of MD5 is particularly weak, given that it was considered to be broken in the cryptographic sense long before the paper was published. Furthermore, it is the Soundex encoding that is being hashed, not the actual name. As a result the input set is even smaller than it would normally be. The reversing of the hash would be trivial. However, this would not lead to recovery of exact names, due to the Soundex encoding being lossy, it would only permit recover of the set of names that encode to that Soundex. 

The main aim of the scheme appears to be to avoid frequency attacks, by injecting fake names to ensure all Soundex codes have the same frequency. This may defeat frequency attacks, but Soundex codes will be easily recoverable, making the protection against frequency attacks a moot point. The concept of injecting fake or duplicate values in order to smooth a frequency distribution is a sound one, though implemented in an insecure protocol in this instance. Such an approach could be considered in a proposed solution, in order to help protect against frequency attacks, and therefore warrants further consideration. 

\subsection{Secure Multi-Party Computation}\label{sec:lit:smc}
Secure Multi-Party Computation is a large research field in its own right, which is beyond the scope of what we can review in this document. In general, SMC approaches offer stronger security guarantees, often based on the underlying mathematical guarantees of the cryptographic primitives.  The key idea is to distribute the computation among two or more participants so that privacy and correctness can fail only if the attacker compromises more than a threshold of participants.  For example, when there are two participants it is often possible to prove that one of them receives only the properly-linked information they need, while no information is leaked to the other.  This sort of model would be very appropriate for ABS.

However, the computational cost of the schemes is often prohibitive if deployed at scale. There is also a significant increase in the capability required for implementation and support of a SMC protocol, due to the complex cryptography involved. Furthermore, evaluating that the implementation is a faithful replication of the design is non-trivial, and failure to comply with the design can have disastrous effects on the security of the protocol. Nonetheless, SMC is undoubtedly the future of privacy preserving record linkage. It will allow two or more parties to link, share and evaluate data in a manner that does not leak information or break privacy. As such, it is worth consideration if only in terms of longer term plans. 

The usage of SMC in privacy preserving data processing is not new. As early as 2001 Du \etal~\cite{du2001protocols} proposed using SMC as way of securely retrieving data from the a database, in a manner in which the database holder does not learn the contents of the query and the querying party does not learn the contents of the database. Their approach used a number of techniques, including Oblivious Transfer. The scheme is quite old now, and more efficient approaches have been presented. However, it should give an indication of the length of time the problem has been studied for, and is still not solved at scale, to provide an indication that it could be several years before a suitable solution is available. 

More recent proposals have attempted to resolve the scaling issue by blocking the dataset prior to performing the expensive SMC protocol. Inan \etal~\cite{inan2010private} propose using Differential Privacy to permit a joint blocking process to take place, whilst limiting information leakage. Following the blocking a SMC protocol is run to calculate the matches. Whilst it undoubtedly improves the efficiency of the overall protocol, the implementation complexity would be significant. 

An area which has driven a large amount of work in privacy preserving matching is in the area of DNA matching. As DNA databases grow the demand for efficient privacy preserving matching also grows. Atallah \etal~\cite{atallah2003secure} propose using homomorphic cryptography to calculate an edit distance between DNA sequences. Such an approach is certainly worth further consideration, with the caveat of it having high complexity of implementation. It should also be noted that whenever encryption of the identifiers occurs directly there is always a decryption risk. The protocols proposed will also be based on multi-party settings, which will require re-analysing in a setting where there is a single entity, as in our context. 

Hall and Fienberg~\cite{hall2010privacy} present a review of Privacy Preserving Record Linkage literature. The focus of the paper is on secure multi-party computation and the remaining challenges. The paper initially looks at existing approaches that are based on hashes and HMACs and highlights that such schemes are insecure. The authors also highlight the different security models, and the limitations of requiring a trusted third party. A number of SMC approaches are discussed, with the overarching conclusion being that they remain computationally demanding, and possibly infeasible when deployed at scale. The authors draw the conclusion that most of the SMC-based protocols rely on either secure set intersection or inner products, and therefore these areas warrant further research to try to achieve the efficiency gains required to make the approaches feasible at scale. The paper discusses the security risks of revealing intermediate values, for example, in the form of similarity scores. It is this very problem that we discuss in relation to ONS~\cite{ONSM9} approach in \cite{culnane2017vulnerabilities}. The authors touch on the concept of thresholded cryptography and how it can play a part in ensuring that intermediate values, like similarity scores, are not revealed and how it can be used to enforce that a series of operations are performed on the cipher texts, to ensure the final decrypted output does not reveal any unintended information.

If a homomorphic approach were to be considered it would require further research into the most appropriate string comparison algorithm. The limitation of operations that can be performed homomorphically will reduce the applicable string comparisons, often to simple edit distance calculations. Yancey~\cite{yancey2005evaluating} provides a comparison of different string comparators for matching in the context of the US Census, which would require further consideration when proposing a homomorphic solution. 

Some schemes~\cite{vatsalan2011efficient} rely on secure set intersection protocols, without providing explicit details. Such protocols are potentially useful when comparing anything from Bloom filters to similarity tables. However, they are often dependent on having two genuinely distinct parties, who both have access to their underlying plaintext data. The efficiency of such protocols remains a problem, which makes them largely unsuitable for our context.

\subsection{Differential Privacy}\label{sec:lit:diffpriv}
\begin{sloppypar}
Differential privacy~\cite{dwork2006calibrating} represents a formal framework for preserving privacy of datasets when releasing statistics. Releases could be scalars or vectors~\cite{dwork2006calibrating}, marginal tables~\cite{barak2007privacy}, models~\cite{chaudhuri2009privacy}, or synthetic datasets~\cite{blum2013learning}. The privacy model of differential privacy is one of untrusted parties receiving the release, where it is assumed that the processing of the data into the release is performed by the data owner. As such it can complement models of untrusted/shared computation. The model of the receiving party is adversarial, which is why the framework has received significant attention recently over earlier frameworks such as $k$-anonymity. For more on the framework refer to the book~\cite{dwork2014algorithmic}. For the task of PPRL on names (with plaintext deleted), for a single ABS party, releasing the unperturbed plaintext linked dataset, the framework is not immediately applicable.
\end{sloppypar}

In addition to Inan \etal~\cite{inan2010private} using differential privacy, Abowd~\cite{AbowdNewton} presented a seminar on Privacy Protection for Statistical Agencies that discussed differential privacy. The seminar did not cover privacy preserving record linkage, but did discuss Statistical Disclosure Limitation within the context of statistical agencies. In particular it discussed evaluating the Marginal Social Cost vs. Marginal Social benefit, and how different research fields assume a different trade-off point. The seminar highlights some important issues that are relevant to the more general setting. In particular that most SDL schemes were conceived when the public sphere was data starved, and as such, the statistical agency could largely determine what information crossed their firewall into the public sphere. In contrast, today, the opposite is true, big data now exists outside the statistical agencies, and to a large extent, not all data held by statistical agencies would even be considered big data. This changes the data release environment and requires different approaches, given both the quantity and lack of control over external data releases. 

Abowd~\cite{AbowdNewton} advocated the public release of documentation to achieve an open and auditable approach, as well highlighting the need for close collaboration between those formally modelling the approaches and the actual implementers. Such points are directly relevant to our context. 

\subsection{UK Office of National Statistics}\label{sec:lit:ons}
The approach proposed by the UK Office of National Statistics in the set of reports ``Beyond 2011'' provides a detailed overview of their research and methods in the area of privacy preserving record linkage. It is applicable to the ABS given the broad similarity of the two organisations. However, there are significant differences in the objectives of the two organisations. First, the threat models are significantly different: the ONS is attempting to protect privacy in the presence of malicious external researchers (albeit extremely restricted researchers), whilst the ABS is attempting to protect privacy from internal actors. Second, the ONS is not constrained by a requirement to delete names and addresses, it is permitted to keep them and is potentially reliant on them when performing some of the linking processes (even in the case of the ``hashed" linking). As such, the ONS approach would not be compliant with the ABS requirement for non-recoverable storage of names and addresses.

The ONS has successfully deployed an HMAC-based linkage identifier using subsets of attributes, like the one described in Option~3.  This seems to work well.  However, while studying for this report we identified weaknesses in other aspects of their methods, which have been reported to them and were published separately \cite{culnane2017vulnerabilities}.



\end{document}